\documentstyle[12pt]{article}
\input psfig
\font\ninept=cmr9
\font\tenpt=cmr10
\def\etal{{\it et~al.}}				
\def\ie{{{\it i.e.},\ }}			
\def\cf{{{\it cf.},\/} }			
\def\CDOT{{$\cdot$} }				
\def\GB140{{GB 140$'$}}
\def\amin{{^{\rm \prime}}}
\def\AMIN{{$^{\rm \prime}$}}

\def\msun{{\,{\rm M}_\odot}}
\def\MSUN{{${\rm M}_\odot$}}
\def\lsun{{\,{\rm L}_\odot}}
\def\LSUN{{${\rm L}_\odot$}}

\def\kms{{\,km\ s^{-1}}}
\def\KMS{{$\rm km\ s^{-1}$}}

\def\HI{{\mbox{\normalsize H\thinspace\footnotesize I}}}

\def\MHI{{$\rm M_{HI}$}}
\def\mhi{ {\rm M_{\ninept HI}} }

\def\MHILB{{$\rm M_{HI}/L_B$}}
\def\MHILFIR{{$\rm M_{HI}/L_{FIR}$}}

\def\LFIR{{$\rm L_{FIR}$}}
\def\LFIRLB{{$\rm L_{FIR}/L_B$}}
\begin{document}
\begin{flushleft}
\large
\centerline {\bf A Neutral Hydrogen Survey of Polar-Ring Galaxies:}
\vspace{0.1in}
\centerline {\bf I.~~Green Bank Observations of the Northern Sample}

\vspace{.75cm}

\normalsize
\centerline {O.-G.~Richter}
\centerline {\it  Hamburger Sternwarte, Gojensbergsweg 112, 2050 Hamburg 80, 
Germany}

~
~
\centerline {P.\ D.~Sackett}
\centerline {\it  Institute for Advanced Study, Princeton, NJ~~08540}

~
~
\centerline {L.\ S.~Sparke}
\centerline {\it  University of Wisconsin, 475 N.~Charter St, 
Madison, WI~~53706}

\vspace{1.0in}
\baselineskip=1\normalbaselineskip

\end{flushleft}

\large
\centerline{\bf Abstract}
\normalsize
\medskip

We present the results of a neutral hydrogen survey 
conducted with the Green Bank 140-foot radio telescope 
of 47 northern objects in the polar-ring galaxy atlas of Whitmore \etal\ 
(1990). 
We detected 39 of these above our 
detection limit of 1.7 \hbox{Jy\CDOT\KMS}; the average measured flux of  
21 Jy\CDOT\KMS\ corresponds to an average neutral 
hydrogen mass of $\rm 5.3 \times 10^9$ \MSUN\ for 
a Hubble constant of $\rm H_0 = 75$ \KMS \ Mpc$^{-1}$.
For the polar-ring galaxies in our sample that have also been observed
with radio arrays, we find that the 21\AMIN\ (FWHM) Green Bank beam often
includes much more flux than found by the synthesis instruments for the polar
rings alone; some of these galaxies are known to have gas-rich companions.
We compare the neutral hydrogen content of the sample to the blue 
luminosity and IRAS fluxes.  The \HI-to-blue-light ratios of the 
confirmed and probable polar rings are around unity in solar units, 
indicating that polar ring galaxies (or their environments) are as gas-rich 
as typical irregular galaxies. 
For their blue luminosity, the confirmed polar rings are underluminous in the
far-infrared, as compared with the rest of the sample.
They are also FIR-underluminous for their \HI\ masses, which suggests that 
most of 
the gas in the ring may be in stable orbits, rather than flowing inward to
trigger star formation in the central galaxy.  
The more disordered class of `related objects,' 
which includes a number of obvious mergers, is highly luminous in the 
far-infrared.

\vspace{0.75in}
\noindent Submitted to {\it The Astronomical Journal\/}.

\vfil\eject

\section{Introduction}
\medskip

A polar-ring system consists of a flattened galaxy with an outer ring of gas,
dust, and stars rotating in a plane approximately orthogonal to the central
galaxy.  The central object is an elliptical or S0 galaxy and 
thus generally devoid of gas and dust.  
Polar rings probably represent merger products, and their study
may give us valuable clues about the process and frequency of merging
(Schweizer \etal, 1983, hereafter SWR).  
In addition, measurements of rotation in
the two nearly-perpendicular planes of the ring and galaxy provide one of the
few probes of the three-dimensional shape of galactic gravitational
potentials, and hence the shape of the dark matter halos 
(SWR, Whitmore \etal, 1987, hereafter WMS, and 
Sackett \& Sparke, 1990, hereafter SS).  
Until recently, only about 20 polar-ring candidates were 
listed in the literature (SWR); 
but a new photometric atlas of polar-ring galaxies (Whitmore \etal, 1990; 
hereafter PRC) now substantially increases the number of 
known polar-ring galaxies and candidates to over 100. 

All but one of the few polar-ring galaxies that have been mapped 
in the 21cm line of neutral hydrogen contain substantial gas,
often a few billion solar masses
(assuming a Hubble constant $\rm H_0 = 75$ \KMS \ Mpc$^{-1}$ used 
throughout this paper),
which is morphologically associated 
with the optical ring and shares similar kinematics (Shane 1980,
Schechter \etal, 1984, and van Gorkom \etal, 1987).   The combination of 
this large gas content and relatively low blue luminosities give
polar-ring galaxies gas-to-light ratios \MHILB\ near unity
(Sackett 1991), even larger than those characteristic of spiral galaxies  
(Giovanelli \& Haynes 1988). 

We undertook the Green Bank 140-foot neutral hydrogen survey described
here primarily to identify a larger 
sample of polar-ring systems rich
enough in \HI\ to be suitable for subsequent high-resolution mapping in
the 21cm line.  
Such mapping measures the distribution and the velocity field of gas in
the rings, both of which are required for an accurate determination of
the shape of the dark matter halo in these systems (\cf SS). 
Together with optical absorption-line studies of the central galaxies,
rotation in the ring gas can determine whether morphological candidates 
are true kinematical polar rings. 
Furthermore, knowledge of 
the ring mass is necessary in order to assess the stability of the ring 
against differential precession, an important consideration in dating 
the ring and hence, presumably, measuring the time since its formation.  
Sparke (1986) has shown that self-gravitating polar rings can remain
stable in a slightly non-polar configuration with a gentle warping
toward the pole, but only if the total ring mass exceeds a lower limit
set by the gravitational potential of the central galaxy.
Since the neutral hydrogen mass in polar rings may 
constitute a substantial fraction of the total ring mass, 
measuring the \HI\ content of a larger sample of polar rings should
show whether self-gravity is generally important in these systems.  
Additionally, our Green Bank data provide the first 
measure of the redshift of two galaxies in our sample.

\section{The Polar-Ring Galaxy Sample}
\medskip

The objects in the 
PRC are divided into four main categories: \smallskip

A:  Kinematically-confirmed polar-ring galaxies (6 objects) \\
\indent B:  Good candidates for polar-ring galaxies (27 objects) \\
\indent C:  Possible candidates for polar-ring galaxies (73 objects) \\
\indent D:  Systems possibly related to polar-ring galaxies (51 objects)\\

The search procedure used to generate the catalog of these relatively rare 
objects was neither systematic nor complete, and the typing of objects into 
the four main categories listed above was necessarily subjective.  
The identification of polar rings is hampered by the fact that their 
appearance is strongly affected by ring brightness and viewing angle. 
Nevertheless, the PRC provides the current best compilation of 
galaxies with an optical component that is severely inclined to the 
plane of the central object.  

Our initial source list consisted of all objects in the PRC with
declinations north of $-45^\circ$ satisfying one or more of the following
criteria:  (1)~known redshift, or (2)~measured blue magnitude $\rm m_B$ 
brighter than 15.5, or (3)~member of PRC category A or B (see above).  
Of the resulting 99 objects, 70 had known redshifts, and 15 had no
measured redshifts but were brighter than $\rm{m_B = 15.5}$.  The
remaining 14 objects were members of PRC category B with unknown
redshifts.

We then checked the electronically-available \HI\ catalog of Huchtmeier and
Richter (1989) 
for previously measured fluxes and 
eliminated several well-observed galaxies from our source list, although 
we did reobserve some of these galaxies in order  
to compare fluxes within the relatively large beam of the Green Bank 140-foot  
telescope with those from other telescopes with smaller beams.  Finally, 
we discarded all galaxies with known redshifts above 8000 \KMS. 
(One target was later found to be listed as slightly fainter than 
$\rm m_B = 15.5$ in newer catalogs, and, prior to the observing run, one 
B category galaxy was measured to have a redshift in excess 8000 \KMS; 
we retained both galaxies in our sample.)   
The resulting source list consisted of 58 objects, 11 of which we were 
unable to observe because of strong interference from the Sun, leaving a 
working sample of 47 objects.

With the Parkes 64m telescope, we have also carried out similar
observations of polar-ring objects with declinations below the southern
limits of Green Bank; those results will be reported in a subsequent
paper.  
Observations of many of the detected galaxies in our northern sample and 
others that we were unable to observe due to interference have been made  
with the Effelsberg 100m radio telescope (Huchtmeier \etal, in preparation). 

\vspace*{.2in}

\section{The HI Observations and Results}
\medskip

The data were obtained during a 9-day observing run in August and September
of 1990 at the 
NRAO\footnote{The National Radio Astronomy Observatory is operated
by AUI, Inc.\ under a cooperative agreement with the National Science
Foundation} 140-foot radio telescope (\GB140) in Green Bank, WV.  
The receiver was the one used previously
at the former 300-foot telescope.  The system noise temperature varied with
elevation:  at best it was 18K; at worst it was about 30K for targets 
close to the southern declination limit of $-45^\circ$.  
The backend was the Mark~IV
autocorrelator with 1024 channels, which was split into four banks of 256
channels each with an IF bandwidth of 10~MHz.  Both polarizations were fed
into two autocorrelator banks 
set to cover the same frequency band in order 
to recover about 8\% of the quantization
loss during the autocorrelation.  The velocity resolution was thus 8~\KMS.
Most ON-source scans were 10 minutes long; 
only the strongest nearby \HI\ sources (\ie NGC~660 and all secondary 
\HI-line calibrators) were observed for shorter
periods.  The ON-scans followed corresponding 
OFF-source scans of equal duration taken at positions generally 
10.5 minutes earlier in right ascension.  
Where the relative positions of program galaxies were favorable, we were able  
to increase our efficiency by using a single 
OFF-source scan for more than one galaxy.
At least two ON/OFF scan pairs were obtained for all
galaxies.  
Due to strong interference from the Sun between 9$^h$ 30$^m$ and 11$^h$ 30$^m$,
some objects that were on our original list could not be observed,  
namely those galaxies with PRC designations B-9 (UGC~5119), C-30 (UGC~5101), 
C-32 (IC~575), C-33 (ESO~500-G41), C-37 (UGC~6182), D-08 (ESO~305-G21),
D-12 (UGC~4892), D-14 (UGC~5485), D-16 (NGC~3406), D-23 (NGC~4753), 
and D-24 (AM~1257-222).  

Figure~1 shows plots of the spectra obtained, after subtraction of a
polynomial baseline fit; the fit was generally 4th order, though
higher-order fits were used where the signal was strong. 
Antenna temperature in mK (where 1 Jy = 0.277K) is plotted on the 
vertical scale in Fig.~1; sources in the OFF-beam appear as 
`absorption' features in the plotted spectra.  
In 39 of the 47 galaxies that we observed, \HI\ flux was detected above the 
3-$\sigma$ r.m.s.\ noise level in at least one channel,   
in profiles that were reproducible in subsamples of equal duration.
For those galaxies observed with the longest integration times, 
this corresponds to a flux detection limit of 1.7 Jy\CDOT\KMS\ for a 
flat-topped line with a 250 \KMS\ linewidth, or a lower limit 
on the neutral hydrogen mass of about $1 \times 10^9 \msun$ of \HI\ at 50 Mpc.  
The average detected flux for this sample of 39 galaxies at the \GB140\ 
was 21 Jy\CDOT\KMS, corresponding to an average detected \HI\ 
mass of $\rm 5.3 \times 10^9$ \MSUN. 
The average linewidth of our detected galaxies is 260 \KMS.
Due to an unreliable optical redshift, one target, NGC~5122, 
was observed at an incorrect radial velocity; subsequently, 
it has been detected in the 21cm line at the VLA 
(Cox, \etal, 1994, in preparation).  
\vfil\eject

~

\vfil\eject

~

\vfil\eject

~
\vspace{2.6in}

\medskip
\centerline{\psfig{figure=HIfig1.ps,height=4.5truein}}

\medskip
{\tenpt 
   
\noindent {\bf Fig.~1~~} 
Spectra obtained at the Green Bank 140-foot (\GB140) 
radio telescope.  
Antenna temperature is shown in mK on the vertical scale; the horizontal 
scale is displayed in \KMS\ along the bottom of each spectra.  
Each spectrum is the result of subtracting 
one or more ON-source/OFF-source pairs and has been baseline-subtracted and 
Hanning-smoothed (except UGC~5600 and NGC~660, which have not been smoothed).
See Table~1 and the Appendix for details about individual objects.
}

\vfil\eject

Table~1 gives the results of our \HI\ observations and related optical data.
In columns 1 and 2 we give the PRC number of the galaxy 
and its conventional name; 
columns 3 and 4 the B1950.0 coordinates; 
columns 5 and 6 the optically-measured heliocentric radial 
velocity and its bibliographic source; 
columns 7 and 8 the total blue magnitude ${\rm B_T}$ 
of each galaxy with its source; columns 9 and 10 the logarithm 
of the optical major axis diameter and its source; column 11 
the newly-determined \HI\ heliocentric radial velocity;
columns 12 and 13 the integrated flux and its mean error; 
column 14 the \HI\ linewidth at 
20\% of the peak; and finally, columns 15 and 16 the peak flux density 
and r.m.s.\ flux density.

For non-detections, column 11 gives the velocity range searched, and the 
total flux integral and peak flux density represent upper limits.  
The upper limit on the peak flux density in column 15 is taken to 
be 3 times the r.m.s.\ noise; this is integrated over a `top-hat' profile 
of width 250 \KMS\ to obtain the upper limit on the 
flux integral given in column 12.  
If the undetected systems are similar to those that we did detect,  
these upper limits are conservative for most objects since 
the HI profiles are more sharply peaked than the flat-topped box we have 
assumed.

Signals detected in the OFF-scan for a given galaxy are reported in 
the subsequent line in Table~1, and are labeled with `OFF.' 
To properly interpret the data in Table~1 for individual
galaxies, and for comments about the individual spectra, the reader is
referred to the notes in the Appendix.  
Optical photographs of the galaxies in our sample can be found in
the polar-ring atlas of Whitmore \etal\ (PRC, 1990).

A few qualifications must be made.  The optical data do not form a  
homogeneous set; where possible we have used information from the 
Third Reference Catalog (de~Vaucouleurs \etal\ 1991, hereafter RC3), but in 
general have had to draw from a variety of sources; the key to these sources 
is given in the caption to Table~1.  If detailed  
optical studies have been performed on particular galaxies, we have used 
values from these works rather than those from the RC3.  Since polar rings 
are often faint, the catalogued diameter measurements will sometimes 
refer to the size of the ring and sometimes to the size of the central 
host galaxy.  The blue magnitudes have been collected from a variety of
sources and cannot be assumed to be on the same scale.  
In some cases the total blue magnitude, $\rm B_T$ was unavailable, and the 
apparent blue magnitude, $\rm m_B$, has been reported instead.

In Table 2 we present infrared data collected for the galaxies in our sample.
Columns 1 and 2 give the PRC identification and the conventional name 
for the galaxy; column 3 identifies the bibliographic source 
of the IRAS flux densities as either the Faint Source Catalog-Version 2 (FSC2), 
co-added scans, or an integrated infrared map; 
columns 4, 5, 6, and 7 give flux densities in the 12-, 25-, 60- 
and 100-micron bands of IRAS, respectively; columns 8, 9, 10, and 11 
the corresponding flux qualities; 
column 12 the logarithm of the total 
far-infrared flux; and columns 13, 14 and 15 the 
12/25, 25/60 and 60/100 infrared colors.  Only flux densities with quality  
2 or better were used for the calculation of the infrared 
colors.  For flux 
\break

~

\vfil\eject

~

\vfil\eject

~

\vfil\eject

~

\vfil\eject


\noindent \hbox{densities} determined from co-added scans, 
columns 8-11 indicate whether the source was resolved in each of 
corresponding IRAS bands. 
 
All IRAS fluxes were retrieved electronically from NED\footnote{
The NASA/IPAC Extragalactic Database (NED) is operated by the Jet 
Propulsion Laboratory, California Institute of Technology, under contract 
with the National Aeronautics and Space Administration.}.  For inclusion 
in NED, FSC2 sources must pass criteria that are designed to 
filter out most stars, including a required detection in the 60 micron band 
(\ie flux quality greater than one) greater than one-half 
the 25 micron flux.  Consequently, it is possible that some identifications 
have been missed, and statistical statements about the 25/60 color of our 
sample should be avoided. 
The FSC2 is signal-to-noise ratio limited 
and, because of reliability requirements,   
does not achieve the full sensitivity of the Faint Source Survey (FSS) plates, 
but is about a factor of 2 to 2.5 times more sensitive than the Point Source 
Catalog (PSC).  Typically, the FSS contains sources with flux densities 
above 0.2 Jy at 12, 25, and 60 microns, and above 1.0 Jy at 100 microns, 
though due to uncertain subtraction of infrared cirrus, especially at 100$\mu$, precise limits will vary with the location of the source on the sky.  

The far-infrared flux, ${\rm F_{\rm FIR}}$, is taken here to be given by  
$${\rm F_{\rm FIR}} = 1.26 \times 10^{-14} \ (2.58 \ {\it f}_{\rm 60} 
+ {\it f}_{\rm 100}) ~~~{\rm W \cdot m^{-2}} \eqno(1)$$

\noindent where ${\it f}_{\rm 60}$ is the flux density at 60$\mu$ and 
${\it f}_{\rm 100}$ is the flux density at 100$\mu$ in Jy.  This 
prescription is used by IPAC (Lonsdale \etal\ 1989) 
to represent the total far-infrared flux 
that would be emitted by a thermal source with the indicated flux densities 
through a perfect bandpass of 80 microns centered on 82.5$\mu$.   This 
representation is valid for dust temperatures between 20K and 80K and 
dust emissivity indices~$n$ (emissivity $\propto \nu^{+n}$) from 0 to 2, which 
are thought to be the ranges relevant to galaxies.  
Color corrections have not been applied.  
If the flux density in one of these two bands has flux quality equal to one, 
and therefore represents only an upper limit, the corresponding upper limit 
on $\log{[\rm F_{\rm FIR}]}$ is given in column 12.  
The infrared colors 
given in columns 13, 14, and 15 are defined here as 
$\log{[{\it f}_{12}/{\it f}_{25}]}$, 
$\log{[{\it f}_{25}/{\it f}_{60}]}$, and 
$\log{[{\it f}_{60}/{\it f}_{100}]}$ respectively.

Table 3 contains quantities derived from the basic optical, radio, and 
infrared parameters given in Tables 1 and 2.  Columns 1 and 2 give the 
PRC number and conventional name for the galaxy; 
column 3 an abbreviation for the radio telescope used; column 4 
the computed distance to the source; column 5 the \HI\ mass 
for the indicated observation; columns 6 and 7 the 
blue and far-infrared luminosities; and columns 8, 9, and 10 
a measure of the ratios of luminosities in the blue, far-infrared, 
and the 21cm line.

Distances in Table 3 are derived from radial velocities 
referred to the centroid of the Local Group according to the prescription 
of the Revised Shapley-Ames Catalog 
\vfil\eject

~
\vfil\eject

~

\vfil\eject


\noindent (Sandage \& Tammann 1987; RSA), namely
$$v_{\rm LG} = v_{\rm helio} + [\ -79 \cos \ell \cos b + 296 \sin \ell \cos b 
- 36 \sin b\ ] ~~~~ {\rm km \cdot s^{-1} }\eqno(2)$$
%
%
Wherever possible, the newly-determined radial velocities from this work 
are used to compute the distance; 
for non-detections, optical velocities from Table~1 are used instead.  
The \HI\ mass from our Green Bank observations 
is calculated directly from the flux integral and the
distances derived from these observations, and does {\it not\/} include any
correction factors, such as for the source (over-)filling 
the beam.  That is, we use simply 
$$\mhi \ {\rm = \ 2.356 \times 10^{5} \ D_{\rm Mpc}^2 \ 
\int S_{\nu, \ Jy} \cdot\ dv_{\kms} }~~~~~~\msun \eqno(3)$$

For the kinematically-confirmed polar-ring galaxies, which are given the 
`A' classification in the PRC, we have searched the literature for 
other \HI\ measurements with which to compare our fluxes.  
Wherever possible, the \MHI\ quoted from these works has been converted 
to our Hubble constant and velocity zero-point for inclusion in Table~3  
and in the Appendix. 
If the total flux 
integral and heliocentric velocity have been given in the original work, 
we use them together with equations (1) and (2) and our Hubble constant 
to derive a distance and an \HI\ mass.  
If the assumed distance and a heliocentric velocity are given, 
we compute a distance based on our assumptions as described above, 
and then multiply the \HI\ mass in the original work 
by the square of the ratio of these two distances to give the mass in Table~3. 
If only an assumed Hubble constant and heliocentric velocity are 
given in the original work, the original \HI\ mass is multiplied by 
the square of the Hubble constant ratio to obtain the \HI\ mass 
appropriate to our Hubble constant.  
In the cases for which we detected no \HI\ emission, an upper limit on 
the \HI\ mass is calculated from the optically-measured redshift and the 
upper limit on the flux integral listed in Table 1.  

The total blue and far-infrared luminosities are computed 
using the distances given in column 4, and the magnitudes and fluxes 
from Tables 1 and 2.  
The blue magnitudes are corrected for external extinction from dust in  
the Milky Way using the $\rm A_B$ values given in the RC3; 
no correction is made 
for extinction internal to the galaxies themselves.  
The resulting blue luminosity is given in column 6 of Table~3 
relative to the solar blue luminosity, $\rm L_{\rm \odot,B}$, 
while in column 7 the far-infrared luminosity is normalized to the bolometric
luminosity of the Sun, $\rm L_{\rm \odot}$; 
solar quantities are taken from Appendix 1A of Binney and Tremaine (1987). 
For the computation of the ratio of far-infrared to blue luminosities given 
in column 9 of Table~3, the fluxes $\rm F_{\rm FIR}$ and $\rm F_B$ were 
first converted to 
common units of $\rm W\cdot m^{-2}$ so that the resulting ratio is unitless.  

\vspace*{.2in}

\section{Discussion}
\medskip

Our survey indicates that polar-ring galaxies are 
associated with much more neutral hydrogen than is usual for early-type 
galaxies.  Figure~2 shows the \HI\ mass as a function of distance for 
all 39 galaxies of our sample that were detected with the \GB140, and 
five upper limits for systems for which redshifts were independently 
available; the detection limit of 1.7 Jy \CDOT \KMS\ is shown as 
the dashed line.  
In this and all subsequent figures, Kinematically-Confirmed polar rings 
(PRC Category A) are shown as filled circles; Good Candidates (PRC Category B) 
are shown as filled squares; Possible Candidates (PRC Category C) are shown 
as open squares; and Related Objects (PRC Category D) are shown as open stars.


\bigskip
\centerline{\psfig{figure=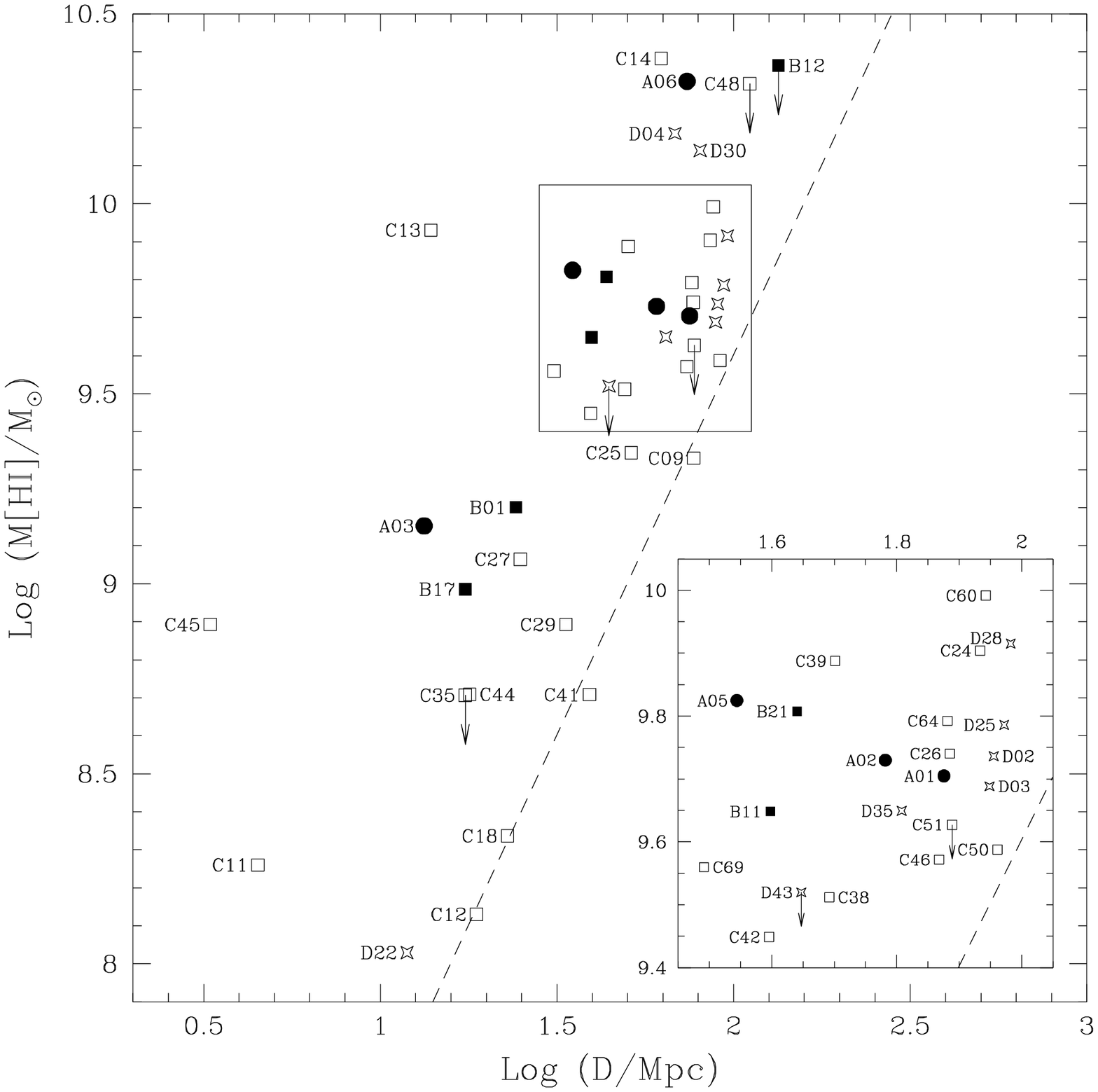,height=4.5truein}}

\medskip
{\tenpt 
   
\noindent {\bf Fig.~2~~} 
The derived \HI\ mass (or upper limits) as a 
function of distance, D, for the galaxies in the \GB140\ sample.  Galaxies 
given the PRC designations A, B, C, and D, are shown as filled circles, 
filled squares, open squares, and open stars, respectively.  The detection 
limit 1.7 Jy \CDOT \KMS\ of the observations is shown as the dashed line.
The insert magnifies the region near D = 51 Mpc and 
\MHI\ = $4.5 \times 10^{9}$ \MSUN, 
the median values of the distance and \HI\ mass for the detected galaxies 
in our sample.
}
\bigskip

The average detected gas mass for our sample 
is $\mhi\ = 5.3 \times 10^9$ \MSUN, 
about twice the level of the S0 galaxies detected in the survey of
Wardle and Knapp (1986).
Objects with \MHI $> 6 \times 10^9 \msun$ include members of all 
four categories of the PRC.  
Based on the results of earlier synthesis mapping, we would expect that the
\HI\ in a polar-ring system is associated with the ring and not with
the central galaxy; consistent with this, our \HI\ profiles fall off 
steeply at the sides and have linewidths typical of galactic rotation, 
as they should if the gas follows stable orbits extending 
well into the flat portion of the rotation curve.  
This is in contrast to the \HI\ profiles in small groups of galaxies, 
where sloping shoulders on the velocity profile are a signature of tidal  
interaction (Gallagher, Knapp \& Faber 1981). 

\bigskip
\centerline{\psfig{figure=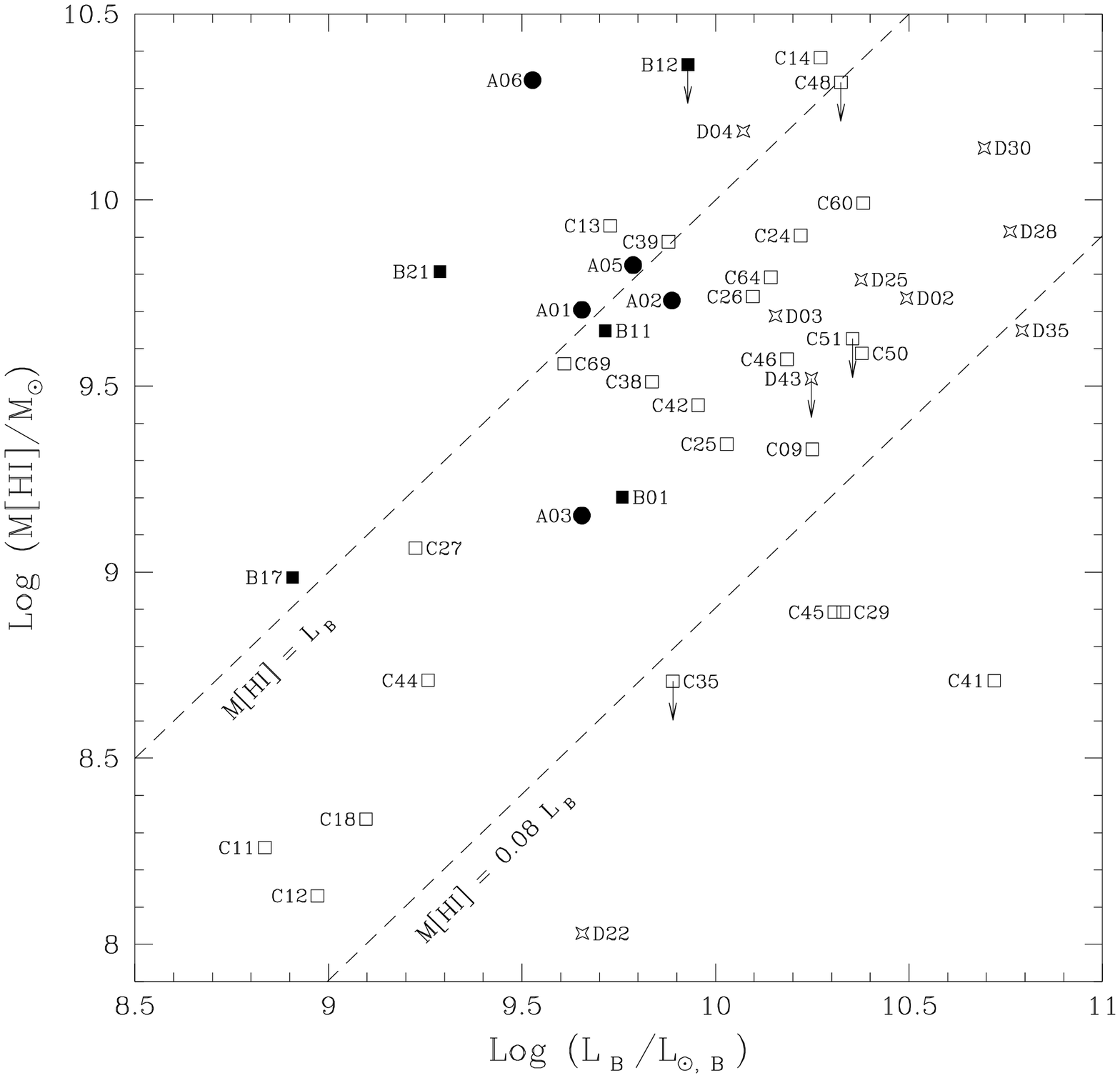,height=4.5truein}}

\medskip
{\tenpt 
   
\noindent {\bf Fig.~3~~} 
The \HI\ mass as a function of blue luminosity for the 
galaxies of the \GB140\ sample; solar units are used.  The upper dashed 
line represents the typical gas-richness of irregular galaxies, \MHILB\ = 1; 
the lower line represents a value more typical of S0 galaxies, \MHILB\ = 0.08.
}
\bigskip

The neutral hydrogen mass is shown as a function of blue 
luminosity in Figure~3; the upper dashed line indicates a gas-to-light ratio 
of \MHILB = 1 in solar units, characteristic of galaxies 
later than type Sc (Giovanelli \& Haynes 1988), and of the most gas-rich
dwarfs in Virgo (Bothun \etal\ 1985).  Objects from PRC Categories A and B 
cluster about this line, while most from Categories C and D 
lie well below it, but still above the mean of 
\MHILB= 0.08 that Wardle and Knapp (1986) derived for their S0 sample.
The `Kinematically Confirmed' polar rings and the `Good Candidates' thus 
show more gas mass per unit blue light than the `Possible Candidates' 
and `Related Objects.'  
Gas-rich Category C and D objects are optically bright, 
while Categories A and B consist of rather faint galaxies, 
all with $\rm L_B \leq 10^{10} \lsun$, that are disproportionately gas-rich 
compared to the rest of the sample.   

\bigskip
\centerline{\psfig{figure=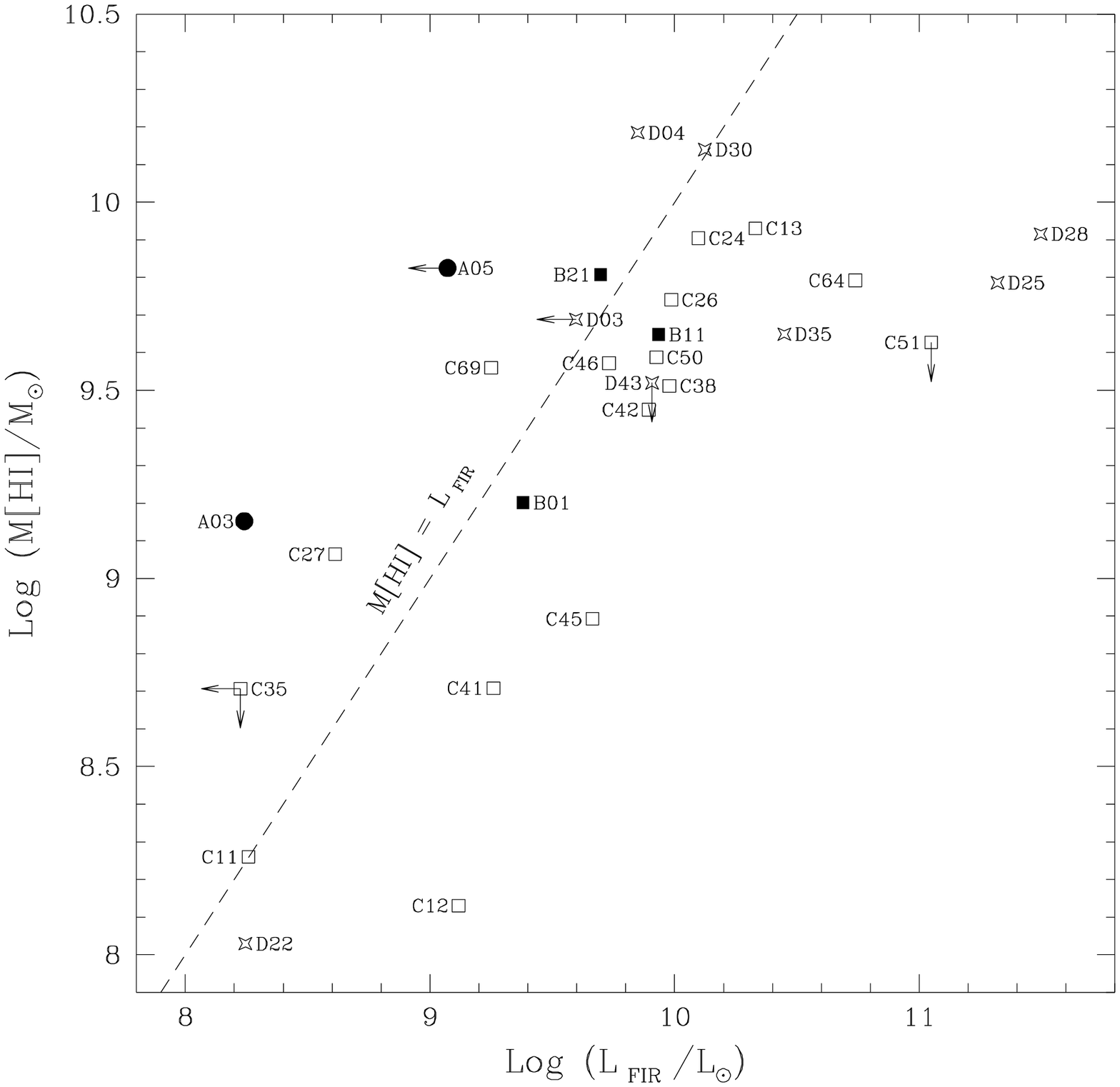,height=4.5truein}}

\medskip
{\tenpt 
   
\noindent {\bf Fig.~4~~} 
The \HI\ mass as a function of far-infrared luminosity in 
solar (bolometric) units for the \GB140\ sample.
}
\bigskip

Figure~4 compares the neutral hydrogen mass with the far-infrared
luminosity, \LFIR, which measures emission from warm dust.  
The Category C and D systems that are most overluminous in the far-infrared 
for their \HI\ mass, compared with the mean of our sample, are 
known interacting and merging galaxies.
The two confirmed polar-ring systems (Category A objects) are the two most 
gas-rich galaxies in the sample for their FIR luminosity.  The three other 
confirmed polar rings in our sample were not detected by IRAS; if we use 
rough upper limits for their 60 and 100 micron fluxes of 0.2 and 1 Jy 
respectively, we find that all three have \MHILFIR $> 1$ in solar units.  
The far-infrared fluxes for the 26 galaxies of our \GB140\ sample 
that were detected in both the 60 and 100 micron bands of IRAS vary over 
more than three orders of magnitude, with a median value of  
\LFIR\ = $8.0 \times 10^9$ \LSUN\ somewhat below the 
median value of \LFIR\ = $1.3 \times 10^{10}$ \LSUN\ derived for a 
sample of interacting galaxies by Bushouse, Lamb \& Werner (1988).  
The median value of \LFIR\ for the 7 Category D galaxies 
is about equal to the median value for the Bushouse \etal\ interacting
sample, while the median of the Category A and B galaxies is about an
order of magnitude lower. 

\bigskip
\centerline{\psfig{figure=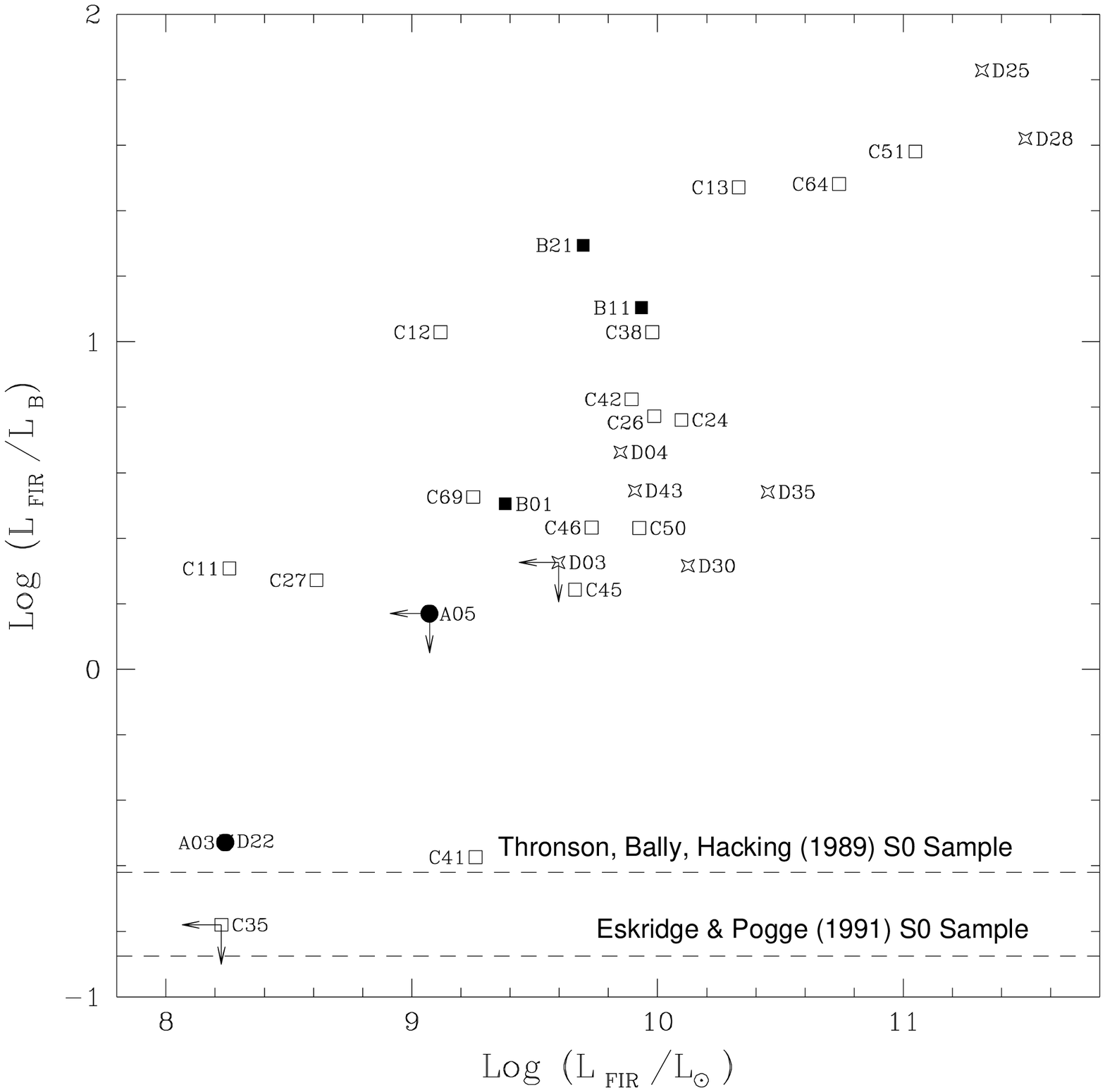,height=4.5truein}}

\medskip
{\tenpt 
   
\noindent {\bf Fig.~5~~} 
The ratio of far-infrared to blue luminosities 
(in W$\cdot$m$^{-2}$) plotted as a function of blue 
luminosity (in solar bolometric units) 
for the galaxies of the \GB140\ sample.  For comparison, the mean values of 
$\log{[\rm L_{\rm FIR}/L_{\rm B}]}$ 
are shown as dashed lines for two S0 samples in the literature.
}
\bigskip

Figure~5 shows the ratio of far-infrared to blue light as a function 
of FIR luminosity for our sample; the luminosities have now been
converted to common units of $\rm W\cdot m^{-2}$.
The median value of \LFIRLB\ for the complete 
\GB140\ sample is 4.1, to be compared with 
5.6 reported by Bushouse \etal\ (1988) for their interacting sample,
and 3.0 for their isolated sample of galaxies.  
Ring galaxies, in which active star formation is seen in a ring that is 
coplanar with the central galaxy, are thought to be the result of a 
head-on collision with a small intruder; these galaxies have a 
larger average far-infrared 
luminosity, \LFIR\ = $2.1 \times 10^{10}$ \LSUN\ (corrected to our H$_0$), 
than our sample, but a considerably smaller \LFIRLB\ = 0.86  
(Appleton \& Struck-Marcell 1987).

Even given their disturbed morphology, it is a little surprising 
that {\it all\/} of the galaxies 
in our polar-ring sample are more FIR-luminous for their blue light 
than the mean value of 
$\log{[\rm L_{\rm FIR}/L_{\rm B}]} = -0.875$ derived by Eskridge and 
Pogge (1991) for a sample of S0 galaxies.  Eskridge and Pogge estimated 
the internal extinction in their sample when calculating $\rm L_B$, and 
performed a survival analysis that took into account upper limits; both would 
tend to lower their mean value relative to our analysis.  Thronson, 
Bally and Hacking (1989) estimated   
$\log{[(2.56 f_{60} + f_{100})/f_{\rm B}]} = 1.5$ for a variety of S0 subtypes 
from the RSA, where ${\it f}_{60}$, ${\it f}_{100}$ and 
${\it f}_{\rm B}$ are flux {\it densities.\/}  
When these are converted to fluxes 
this corresponds to $\log{[\rm L_{\rm FIR}/L_{\rm B}]} = -0.62$, 
considerably below most  
of our sample, but consistent with Eskridge and Pogge's result within the large 
uncertainties of this measurement.   Non-detections by IRAS of three 
Category~A polar rings in our sample places upper limits on their 
$\log{[\rm L_{\rm FIR}/L_{\rm B}]}$ of 0.4 to 0.9.  
The one confirmed polar-ring galaxy (NGC~2685) 
for which IRAS measured flux in both the 60 and 100$\mu$ bands has  
\LFIRLB\ and a 60/100 color similar to that found for early-type galaxies; 
Thronson \etal\ (1989) found the same to be true for their 
sample of shell galaxies, where the structure is also thought to 
represent an `undigested' merger. 

Figures~4 and 5 taken together show that, compared to the rest of our 
sample, the two confirmed polar rings detected by IRAS are severely
underluminous in \LFIR\ for their blue luminosity and their \HI\ mass.  
A possible explanation is that in polar-ring systems the dust resides
mainly in the rings, along with the \HI, while the stars that could
heat it to IRAS-emitting temperatures are mostly in the central galaxy;
much of the dust in these systems may thus be too cool to emit in the
IRAS wavebands. 
The very red 60/100 infrared color of NGC~2685, the reddest in our sample, 
is consistent with primary dust heating from an older population of disk 
stars, rather than from the UV radiation of OB stars 
(Bothun, Lonsdale \& Rice 1989).  
In their study of S0 galaxies, Eskridge and Pogge (1991) find that of 
the objects which are atypically HI rich for their IRAS flux, many of 
those that have been mapped in \HI\ have their gas in outer rings
rather than in the inner disk (van Driel \& van Woerden 1991); this is a
similar situation to polar rings.  The low FIR emission of polar-ring 
galaxies suggests 
that the gas is in stable orbits in the ring, rather than flowing 
rapidly in to the central galaxy, where it would lead to star formation.
As illustrated by Figure~6, the objects of our sample show no obvious 
trend of gas-richness, as measured by \MHILB, with FIR luminosity.  

It is not obvious that all of the \HI\ that we detect is actually in the 
polar-ring galaxies themselves.  
The $21'$ beam of the \GB140\ covers a linear diameter of
300~kpc at a distance of 50~Mpc, which is large enough to include close
companion galaxies; we list nearby companions in the notes. 
For ten galaxies in our sample, radio synthesis maps are available 
from which the mass of gas in the polar ring can be estimated; 
in seven of these, this quantity is more than 30\% below our measured flux.
For five of those seven galaxies, one or more gas-rich
companions were detected in synthesis maps, close enough to contribute 
to the measured Green Bank flux; another has a close neighbor at unknown 
redshift.
Thus the masses that we quote here are only upper limits, and the \HI\ 
mass in the polar-ring galaxies themselves may be only half or a third
as much.  Even so, these are gas-rich galaxies,
in gas-rich environments. 

\bigskip
\centerline{\psfig{figure=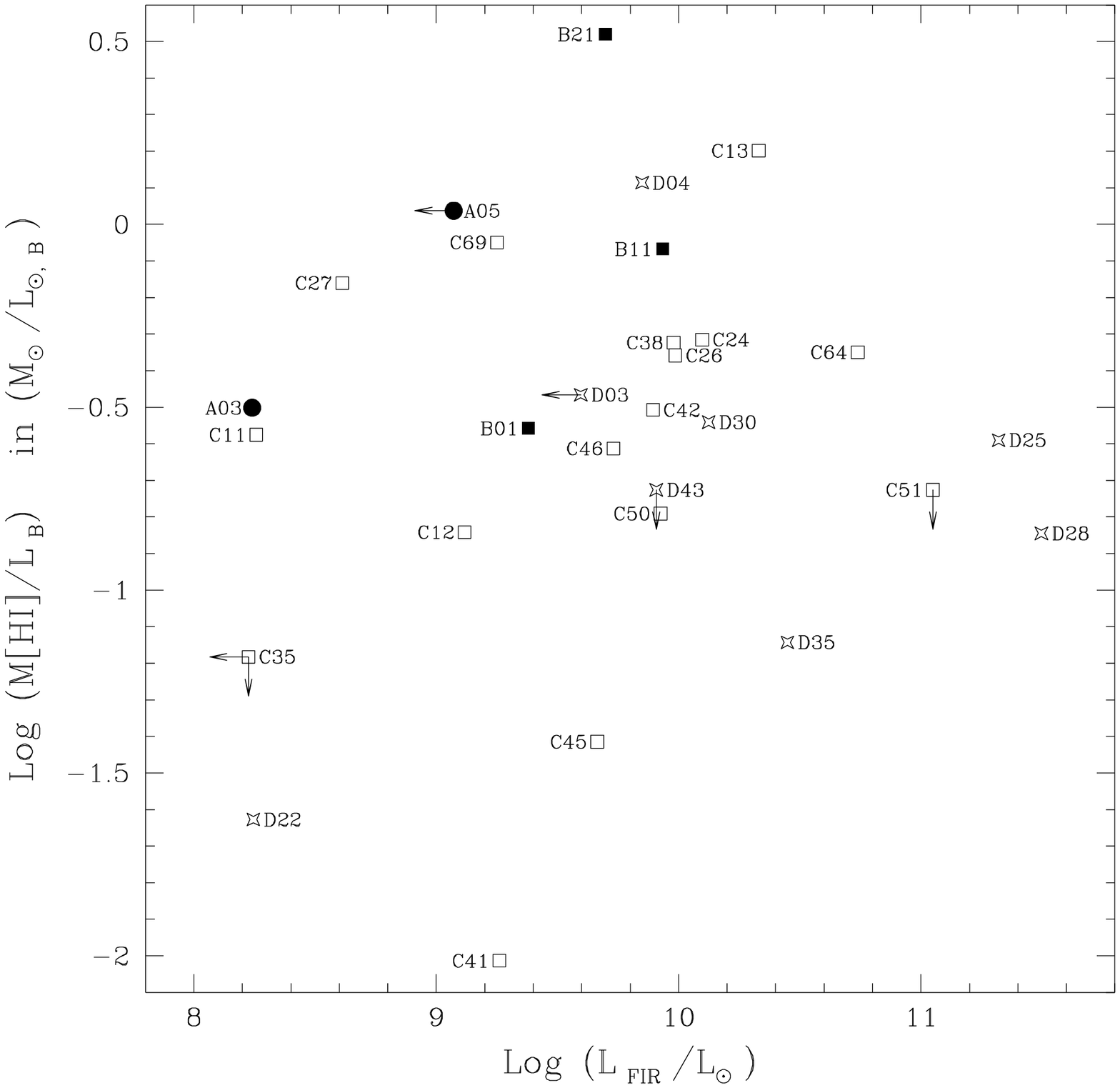,height=4.5truein}}

\medskip
{\tenpt 
   
\noindent {\bf Fig.~6~~} 
The gas-richness, \MHILB, as a function of far-infrared 
luminosity for the \GB140\ sample; no trend is apparent.
}
\bigskip

Polar rings are thought to represent captured material, but 
it is not clear what kind of accretion process would lead to their formation.
Capture of gas from a passing spiral seems 
unlikely, since our observed masses are comparable to the {\it entire\/} 
\HI\ content of a large spiral.  
The suggestion has been made that the rings represent the remains of a
gas-rich dwarf irregular galaxy, which has been captured and torn apart
by differential rotation.  
However, the distribution of \HI\ masses in dwarf irregulars peaks
around $5 \times 10^8 \msun$ (Matthews, Gallagher \& Littleton 1993), a
factor of ten below what we find for our polar ring sample.  
Further, our computed ratios of \HI-mass-to-light include the central
galaxy, which contributes most of the light but (on the basis of
synthesis maps), almost none of the gas.  
Thus the accreted object must have been 
unusually gas-rich, 
but with an optical luminosity that is small 
even compared to the relatively dim central galaxies of polar rings.    
Yet another possibility is that the 
rings result from the delayed infall of primordial or intragroup gas.
Further synthesis observations would establish whether diffuse group gas
has contributed substantially to our measured fluxes,
since extended emission would be resolved out in those maps.  
One might question whether such an ordered structure as a polar ring can 
form out of intragroup gas; but the giant ring in Leo (Schneider \etal\
1989), a 200 kpc diameter structure containing at least $10^9 \msun$
of \HI\ that appears to be in coherent rotation, may be a 
larger-scale example of such a phenomenon.

\vspace*{4.5in}

\noindent {\bf Acknowledgements}
\medskip

It is with pleasure that we thank the telescope operators and other NRAO staff,
especially Ron Maddalena, for their high-quality and cheerful support during
the observing session.  The Green Bank site director G.~Seielstad kindly
accommodated a delay in the observing run.  
We are grateful to 
Michael Strauss and John Huchra for assisting us in our 
search for optical redshifts, to Marion Schmitz for tracking down the 
source of a few anomalies in NED, to Frank Briggs for re-reducing VLA 
archival data of NGC~4650A, 
and to Harley Thronson and Rick Pogge for several lively e-mail exchanges.     
This research has made use of the NASA/IPAC Extragalactic Database (NED).  
Most of this work was done while OGR was an ESA staff member at STScI.  
Work by PDS was supported in part by the National Science
Foundation (AST~92-15485) and the J.~Seward Johnson Charitable Trust.  
LSS was partially supported by NSF grant AST~90-20650, and thanks the 
Institute for Advanced Study for hospitality.

\vfill\eject

\section*{References}
\renewcommand\bibitem{\par\noindent\hangindent=2pc\hangafter=1}

\bibitem  
Appleton, P.N. \& Struck-Marcell, C. 1987, ApJ, 312, 566

\bibitem  
Balkowski, C., Chamaraux, P., and Weliachew, L. 1978, AA, 69, 263

\bibitem
Binney, J.J., \& Tremaine, S. 1987, {\it Galactic Dynamics,\/}  
(Princeton University Press: Princeton)

\bibitem
Bland, J., Taylor, K. \& Atherton, P. 1987, MNRAS, 228, 595

\bibitem 
Bothun G.D., Mould, J.R., Wirth, A. \& Caldwell, N. 1985, AJ, 90, 697

\bibitem
Bothun, G.D., Lonsdale, C.J. \& Rice, W. 1989, ApJ, 341, 129

\bibitem
Bushouse, H.A., Lamb, S.A. \& Werner, M.W. 1988, ApJ, 335, 74

\bibitem 
de Vaucouleurs, G., de Vaucouleurs, A., Corwin, H.G. Jr., Buta, R.J., 
Paturel, G. \& Fouqu\'e, P. 1991, 
{\it Third Reference Catalogue of Bright Galaxies,\/} Univ.\ of Texas
Press, Austin (RC3) 

\bibitem
Eskridge, P.B. \& Pogge, R.W. 1991, AJ, 101, 2056

\bibitem 
Gallagher, J.S., Knapp, G.R. \& Faber, S.M. 1981, AJ, 86, 1781. 

\bibitem
Giovanelli, R. \& Haynes, M.P. 1988, in {\it Galactic and Extragalactic 
Radio Astronomy,\/} 2nd edition, eds. G.L. Verschuur \& K.I. Kellerman
(Springer: New York), p522 

\bibitem
Graham, J. A. 1979, ApJ, 232, 60

\bibitem
Hibbard, J.E., Guhathakurta, P., van~Gorkom, J.H. \& Schweizer F. 1993, 
submitted to AJ

\bibitem
Huchra J.P., Geller, M.J., Clemens, C.M., Tokarz, S.P. \& 
Michel, A. 1992, Bull. C.D.S., 41, 31 (Strasbourg) (ZCAT)

\bibitem
Huchtmeier, W.K., \& Richter, O.-G. 1989, {\it A General Catalog of HI
Observations of Galaxies,\/} Springer-Verlag, New York. 

\bibitem
Jarvis, B.J. 1987, in {\it Structure and Dynamics of Elliptical 
Galaxies,\/} IAU Symposium 127 (Reidel: Dordrecht), p411 

\bibitem
Lauberts, A. 1982, {\it The ESO-Uppsala Survey of the ESO(B) Atlas,\/}  
(ESO: Garching) (ESO)

\bibitem
Lauberts, A., \& Valentijn, E.A. 1989, {\it The Surface Photometry Catalogue 
of the ESO-Uppsala Galaxies,\/} (ESO: Garching) (LV) 

\bibitem
Lonsdale, C.J., Helou, G., Good, J.C., Rice, W. \& Fullmer, L. 1989,
{\it Catalogued Galaxies and Quasars in the IRAS Survey,\/} 
2nd version (JPL: Pasadena) 

\bibitem
Mahon, M. E. 1992, PhD Thesis, U. of Florida, Gainesville.

\bibitem 
Matthews, L.D., Gallagher, J.S. \& Littleton, J.E. 1993 in {\it Massive Stars 
and Their Lives in the Interstellar Medium,\/} eds. J.P. Cassinelli \& E.B. 
Churchwell (ASP: San Francisco), p501 

\bibitem
Mirabel, I.F. \& Sanders, D.B. 1988, ApJ, 335, 104

\bibitem
Nicholson, R. A., Bland-Hawthorne, J., \& Taylor, K. 1992, ApJ, 387, 503

\bibitem
Nilson, P. 1973, {\it Uppsala General Catalogue of Galaxies,\/} 
Nova Acta Regiae Societatis Scientarum Upsaliensis, Ser.\ V:A.\ Vol.\ 1 (UGC) 

\bibitem
Rubin, V.C. 1987 in {\it Dark Matter in the Universe,\/} 
IAU Symposium No.~117, eds. J.~Kormendy and G.~Knapp (Reidel, Dordrecht)

\bibitem
Sackett, P.D. 1991, 
in {\it Warped Disks and Inclined Rings around Galaxies,\/} 
eds. S. Casertano, P.D. Sackett \& F.H. Briggs 
(Cambridge University Press: Cambridge), p73

\bibitem
Sackett, P.D., \& Sparke, L.S. 1990, ApJ, 361, 408 (SS) 

\bibitem
Sandage, A. \& Tammann, G.A. 1987, {\it A Revised Shapley-Ames Catalog of 
Bright Galaxies,\/} Carnegie Institution of Washington, Publication 635, 
Second Edition (RSA) 

\bibitem
Schechter, P.L., Sancisi, R., van~Woerden, H., \& Lynds, C.R. 1984,
MNRAS, 208, 111 

\bibitem 
Schneider, S., \etal\ 1989, AJ, 97, 666

\bibitem
Schweizer, F. 1982, ApJ, 252, 455

\bibitem
Schweizer, F., Whitmore, B.C., \& Rubin, V.C. 1983, AJ, 88, 909 (SWR) 

\bibitem
Seitzer P, \& Schweizer, F. 1990, in {\it Dynamics and Interactions of 
Galaxies,\/} ed R. Wielen (Springer: Heidelberg), p270 

\bibitem
Shane, W.W. 1980, AA, 82, 314 

\bibitem
Sparke, L.S. 1986, MNRAS, 219, 657 

\bibitem
Sulentic, J.W. \& Tifft, W.G. 1973, {\it The Revised New General Catalog of
Nonstellar Astronomical Objects,\/} (U of Arizona Press: Tucson) (RNGC)

\bibitem
Thronson, H.A., Bally, J. \& Hacking, P. 1989, AJ, 97, 363

\bibitem
Tully, R.B. 1988, {\it Catalog of Nearby Bright Galaxies,\/} (Cambridge 
University Press) (NBG)

\bibitem
van~Driel, W. \& van~Woerden, H. 1991, AA, 243, 71

\bibitem
van~Gorkom, J.H., Schechter, P.L., \& Kristian, J. 1987, ApJ, 314, 457 

\bibitem
van~Gorkom, J.H., van der Hulst, J. M., Haschick, A.D. \& Tubbs, A.D., 
1990, AJ, 99, 1781   

\bibitem
Wardle, M. \& Knapp, G.R. 1986, AJ, 91, 23

\bibitem
Whitmore B.C. \& Bell M. 1988, ApJ, 324, 741

\bibitem
Whitmore, B.C., McElroy, D., \& Schweizer, F. 1987, ApJ, 314, 439 (WMS) 

\bibitem
Whitmore, B.C., Lucas, R.A., McElroy, D.B., Steiman-Cameron, T.Y.,
Sackett, P.D., \& Olling, R.P. 1990, AJ, 100, 1489 (PRC) 

\vfil\eject

\begin{center}
\section*{Appendix:  Notes on Individual Galaxies}
\end{center}

\noindent

\begin{description}

\item[A-1~=~A0136-0801] 
Our linewidth of 408 \KMS\ is only 
slightly larger than the 342 \KMS\ measured at the VLA 
(van Gorkom \etal\ 1987), yet we detected $5.1 \times 10^9 \msun$ of \HI\ 
with the 20\AMIN-beam (FWHM) of the Green Bank 140-ft (\GB140), 
while the VLA only sees $1.5 \times 10^9 \msun$.   
NED lists no companions within 10\AMIN; of the two companions within
20\AMIN, NGC~636 is outside the velocity range at 1847 \KMS, and
MCG--01-05-012 is a 15th magnitude galaxy of unknown redshift.  
The similarity of the linewidths suggests that most of the 
measured \HI\ mass is associated with the polar-ring system. 

\item[A-2~=~ESO~415-G26] 
We see almost twice as much gas ($5.4 \times
10^9 \msun$) with a linewidth that is more than 100 \KMS\ larger than 
indicated by previous VLA observations (van~Gorkom \etal\ 1987).  
The VLA measured $3.2 \times 10^9 \msun$ of HI, but missed flux at 
high relative velocities, especially on the blueward side on the line,
due to interference
(J.~van~Gorkom, private communication); this may be responsible for
the `butterfly contours' in the total \HI\ map, the smaller linewidth, and 
higher systemic velocity of the VLA observations.    
NED reports no companions within 10\AMIN\ and only one galaxy within 20\AMIN:
RC3~022508$-$315509, with no listed magnitude but a radial velocity (4530 \KMS) 
that places it within the linewidth of ESO~415-G26.  It is thus
unclear whether all of the \HI\ detected at the \GB140\ is associated
with the polar-ring system. 

\item[A-3~=~NGC~2685]
The \HI\ map made at Westerbork by Shane (1980) shows two dynamical
systems in \HI, one associated with inner optical `helix' that crosses
the main optical disk perpendicularly, the other at large radius at a
position angle close to that of the central galaxy.  
His estimate of the total \HI\ mass is $1.2 \times 10^9$ \MSUN, of which
one-quarter is in the central polar ring; our Green Bank observations
find a mass of $1.4 \times 10^9$ \MSUN\ with nearly the same linewidth of about 300 \KMS.  
New VLA data (Mahon 1992) for this galaxy show $1.6 \times 10^9$ \MSUN\
with about the same linewidth. 
One small, 16th magnitude galaxy lies within 10\AMIN\ of NGC~2685 and two more
faint, small galaxies are within 20\AMIN. 
We identify the negative feature at 1288 \KMS\ in the subtracted spectrum 
of NGC~2685 as an OFF-beam detection of UGC~4549.

\item[A-5~=~NGC~4650A] 
The \GB140\ sees $6.7 \times 10^9 \msun$, 
50\% more flux than the VLA ($4.6 \times 10^9 \msun$; van Gorkom \etal\ 
1987); the linewidths are quite similar.  A preliminary
reduction of the VLA archival data (F.~Briggs, private communication)
shows that the nearby spiral companion, NGC~4650, has a 
significant amount of \HI\ flux.  NGC~4650 is $5'$ away from
NGC~4650A, well inside the \GB140\ beam, and has a radial velocity of 2953
\KMS, \ie it is within 100 \KMS\ of NGC~4650A.  There are two other
companions within 10\AMIN\ and another two within 20\AMIN\ with 
similar redshifts; this galaxy is in a fairly dense environment.

\item[A-6~=~UGC~9796~=~II~Zw~73] 
The \GB140\ sees $2.1 \times 10^{10} \msun$
of gas, nearly five times as much as the Westerbork synthesis
observations reported by Schechter \etal\ (1984) for the polar ring alone; 
the linewidths are quite 
similar, but our profile is very asymmetric and is not double-horned.  
Seven companion galaxies were also detected in the Westerbork map;
five of these lie within the \GB140\ beam, including the near neighbor 
MCG +07-31-049 which is separated by $1.5\amin$ and 20 \KMS\ from UGC~9796.  
The companions are partly responsible for the high gas-to-galaxy-light ratio of 
\MHILB\ = 6.2 of this system.  
The total \HI\ mass integrated from the Westerbork map within 12\AMIN\ is 
$1.6 \times 10^{10} \msun$ (R.~Sancisi, private communication), 
close to our result.  
Although no IRAS flux has been reported for UGC~9796, $60\mu$ flux has been 
reported for its companion, MCG~+07-31-049.

\item[B-1~=~IC~51~=~Arp~230] 
The coordinates for this shell galaxy are listed
incorrectly in the PRC; correct coordinates are given in Table 1.
The \GB140\ detected $1.6 \times 10^9 \msun$ of \HI.
NED lists no optical companions within 30\AMIN.

\item[B-11~=~UGC~5600] 
This galaxy shows optical emission in a large, diffuse 
face-on outer ring as well as in the inner, narrow, and 
apparently edge-on ring that is perpendicular to the position angle 
of the central object.  The total \HI\
mass measured by the \GB140\ is $4.5 \times 10^9 \msun$.  
UGC~5600 is paired with its disturbed companion UGC~5609, which is 
located at very nearly the same radial velocity (2729 \KMS) and 2\AMIN\ 
to the south.  NED reports five other galaxies of unknown redshifts 
within 20\AMIN; all but one are faint.

\item[B-12~=~ESO~503-G17]
Despite the large distance of this galaxy, its very suggestive morphology 
warranted inclusion in our sample.  Together with its optical redshift, 
non-detection with the \GB140\ 
implies a weak upper limit of $\mhi = 2.3 \times 10^{10}$.  
No companions listed by NED within 30\AMIN.

\item[B-16~=~NGC~5122] 
An unreliable optical redshift caused us to miss this galaxy in our 
\GB140\ run. The spectrum at the coordinates of NGC~5122, but 
at a higher, incorrect redshift, is shown in Fig.~1;  
it appears that there is significant flux at 4800 \KMS.   
Subsequent observations at the VLA (Cox \etal, 1994, in preparation) 
has revealed that 
about $5 \times 10^8 \msun$ of gas is associated with the ring of 
NGC~5122 itself, and an equal amount is in a companion (MCG--02-34-045) 
10\AMIN\ and about 100 \KMS\ away.  
NED lists no other companions within 20\AMIN.  The VLA map, along with
optical long-slit spectroscopy of the central body (Jarvis and Sackett,
unpublished), confirms this galaxy as a polar ring. 

\item[B-17~=~UGC~9562~=~II~ZW~71] 
We measure a total \HI\ mass of $9.7 \times 10^8 \msun$.
A previous synthesis observation at Westerbork (Balkowski \etal\ 1978) 
indicated a combined \HI\ mass of $1.1 \times 10^9 \msun$ for II~Zw~71,
its companion galaxy II~Zw~70 (UGC~9660) which is 4\AMIN\ away, 
and an intergalactic \HI\ cloud between them.  
Their \HI\ map indicates that about 
$6.3 \times 10^8 \msun$ of this gas is associated with the ring of II~Zw~71. 
This is consistent with an Arecibo single-dish observation 
(\cf Huchtmeier \& Richter 1989) which gives a 
total flux that is two-thirds of that seen with the \GB140, 
within a beam approximately 3\AMIN\ across.

\item[B-21~=~ESO~603-G21] 
We measure an \HI\ mass of $6.4 \times 10^9 \msun$. 
The galaxy has a prominent dust lane that warps across the ring. 
NED lists only one companion within 20\AMIN, the peculiar spiral 
ESO~603--G20, which is 5\AMIN\ away with a redshift (3171 \KMS) close to
that of the polar-ring galaxy.  It is likely that this companion
contributes part of the flux measured by the \GB140, which may explain
the enormously high gas-to-light ratio of $\rm M_{\rm
HI}/L_{\rm B} = 3.3$. 

\item[B-23~=~A~2330--3751]
Since no redshift is available, the \GB140\ observations place an upper limit 
of about 2.2 Jy\CDOT\KMS\ on the total flux in the 21cm line 
in the search band (1000-8000 \KMS), but do not constrain the total \HI\ mass.  
This galaxy lies 8.5\AMIN\ from the center of the cluster Abell~4015.    

\item[B-27~=~ESO~293-IG17~=~AM~2353-392] 
No redshift is available; our non-detection places an 
upper limit on the total flux in the 21cm line of about 2.5 Jy\CDOT\KMS\ 
in the search band (1000-8000 \KMS).  This galaxy is a member of a projected  
triple on the sky, and NED lists one other optical companion, ESO 293-G15, 
4.4\AMIN\ away. 

\item[C-9~=~NGC~442] 
NED reports no companions within the \GB140\ beam that are at a 
redshift consistent with this marginal detection of 
$2.1 \times 10^9$ \MSUN\ of \HI.

\item[C-11~=~NGC~625~=~ESO~297-G5] 
This nearby galaxy is unambiguously detected with the \GB140\ with $1.8 \times
10^8$ \MSUN\ of \HI\ in a spectrum that also includes \HI\ emission from the 
Galaxy near zero velocity.  There are two galaxies about 11\AMIN\ away:  
MS~0132.5--4151 has a much higher redshift (z = 0.172), while ESO~297--G010 
has a blue magnitude of 15.2 and an unknown redshift.

\item[C-12~=~UGC~1198~=~VII~Zw~3] 
An extremely marginal detection of $1.4\times 10^8$ \MSUN\ is deduced from the 
sharp peak at 1100 \KMS.  A NED search shows no nearby companion.

\item[C-13~=~NGC~660~=~UGC~1201] 
This nearby starburst galaxy 
contains optically-emitting gas mixed with dust in a warped ring that is
highly inclined to the central body. 
The strong double-horned profile seen with the \GB140\ contains 
$8.5 \times 10^9$ \MSUN\ of \HI.   
The nearby UGC~1211 is 16\AMIN\ away with a radial velocity 
that is 1600 \KMS\ higher than that of NGC~660; it is not a 
source of confusion for our observations.  
Previous Arecibo observations (\cf Huchtmeier \& Richter 1989) give a flux
integral of 54 $\rm Jy \cdot \kms$, about one-third of our flux, but since the
optical diameter of this galaxy is about 8\AMIN, the \GB140\ beam is 
well-matched to the size of the system, whereas the 3\AMIN\ 
Arecibo beam would be expected to miss emission.  
New high-resolution VLA observations (Mahon 1992)
show about 80\% of the \GB140\ flux and a slightly 
larger linewidth.  It is likely that all the HI that we detect is in 
fact associated with NGC~660, which makes it quite gas-rich, with an $\rm
M_{\rm HI}/L_{\rm B} = 1.6$.  In the far-infrared, NGC~660 is one of
the more luminous galaxies in our sample. 

\item[C-14~=~NGC~979~=~ESO~246-G23] 
The outer optical ring is faint and smooth; it may be 
a face-on polar ring.  The \GB140\ measured 
an \HI\ mass of \hbox{$2.4 \times 10^{10}$ \MSUN}, with a large linewidth of 
678 \KMS; the profile is single-peaked rather than double-horned.  
To our knowledge, this detection provides the only known 
redshift for this galaxy.  
NED lists one companion within 20\AMIN: ESO~246--G22, which is 
9\AMIN\ away and has an optical radial velocity of 5127 \KMS.  This 
companion may have contributed to the measured flux and large velocity 
spread of the gas.

\item[C-18~=~ESO~358-G20] 
The \GB140\ detects this small galaxy with $2.2 \times 10^8$ \MSUN\ 
of \HI\ in a very asymmetric profile.  
The galaxy lies near a 
faint, distant galaxy cluster, but NED lists no companions 
of similar redshift within 10\AMIN, and within the \GB140\ beam one, 
ESO~358-G25, at 13\AMIN\ away and a radial velocity of 1459 \KMS, which may 
be responsible for some of the flux in the asymmetric line.  

\item[C-24~=~UGC~4261] 
The \GB140\ sees a large \HI\ mass, $8\times 10^9$
\MSUN\ within a slightly asymmetric, but double-horned profile.  
NED lists no companions within 10\AMIN, and only two small 
galaxies within 20\AMIN.  This and the profile shape  
suggest that most of the \HI\ is associated with UGC~4261 itself.  

\item[C-25~=~UGC~4323] 
A total \HI\ mass of $2.2 \times 10^9$ \MSUN\ is detected by the \GB140\ 
in an asymmetric profile.    
Emission in the OFF-beam appears as the `absorption' feature at 4785 \KMS\ 
in the subtracted spectrum.  
NED lists no companions within 10\AMIN; the galaxy pair 
including UGC~4353 is about 15\AMIN\ away with a radial velocity of about 
4300 \KMS, but the \GB140\ sees no significant flux at this velocity.  
Our radio redshift differs from the optical one by about 300 \KMS, 
though our peak flux is much nearer the optical velocity.  

\item[C-26~=~UGC~4332] 
The \GB140\ sees $5.5 \times 10^9$ \MSUN\ of \HI\ in a somewhat asymmetric 
profile.  The extensive dust on the NE side of the galaxy may be related 
to its IRAS flux.  This galaxy is in a cluster 
and has many close neighbors.   
Since the radial velocities (as reported by NED) of the neighbors within 
the \GB140\ beam 
are smaller than 5200 \KMS\ or larger than 6500 \KMS, most of the flux 
detected at 5811 \KMS\ is likely to be associated with UGC~4332, 
despite the fact that the line center differs by about 300 \KMS\ from the 
optical velocity.  An OFF-beam detection can be seen in the spectrum 
at 5134 \KMS.  

\item[C-27~=~UGC~4385] 
Clear detection of $1.2 \times 10^9$ \MSUN\ of \HI\ within a double-horned 
profile.  NED lists no optical companions within the \GB140\ beam.  

\item[C-29~=~NGC~2865] 
Within a double-horned profile, 
the \GB140\ sees $7.8 \times 10^8$ \MSUN\ of \HI\ in this shell galaxy.  
The \MHILB\ ratio of this galaxy is only about 0.04.  
NED lists no companions within the \GB140\ beam. 

\item[C-35~=~NGC~3414] 
Our upper limit of $5.1 \times 10^8 \msun$ with \GB140\  
implies a small gas-to-light ratio of $\rm M_{\rm HI}/L_{\rm B} 
\leq 0.07$.  This galaxy has unusual isophotes and is considered by 
Whitmore and Bell (1988) to be a box or X-galaxy; it has three  
neighbors with separations within 10\AMIN\ and redshifts within 300 \KMS.    

\item[C-38~=~NGC~3934~=~UGC~6841] 
The \GB140\ sees $3.3\times 10^9$ \MSUN\ of \HI.  
The baseline was affected by solar interference.  NED lists three other 
galaxies within 15\AMIN; two are small and quite faint, but the third, 
NGC~3933, is only 3\AMIN\ away with a nearly identical radial velocity, 
and so may be contributing some of the \GB140\ flux.  

\item[C-39~=~NGC~4174] 
The \GB140\ sees $7.7 \times 10^9$ \MSUN.  
This galaxy is in a tight \HI-rich group, Hickson~61;
NED lists 3 galaxies within 4\AMIN, two of which 
have redshifts similar to NGC~4174, and another two  with 
similar redshifts about 20\AMIN\ away.  Thus, some of the \HI\ gas 
detected by the \GB140, including the apparent emission at about 3650 \KMS, 
is probably associated with the group rather than with the polar ring.

\item[C-41~=~IC~3370] 
Although generally considered a box-shaped elliptical, there is 
evidence for cylindrical rotation and X-shaped isophotes (Jarvis 1987, 
Whitmore \& Bell 1988, Seitzer and Schweizer 1990).  
NED lists one other galaxy within the \GB140\ beam and three more just 
outside it, one of which has a redshift nearly identical to that of IC~3370.
Our extremely marginal detection of $5.1 \times 10^8$ \MSUN\ of \HI\ 
is shifted by 300 \KMS\ from the optically-determined redshift of the galaxy. 
We find $\rm M_{\rm HI}/L_{\rm B} < 0.01$, which  
makes this system the most gas-poor galaxy in our sample.  

\item[C-42~=~NGC~4672] 
The \GB140\ sees $2.8 \times 10^9$ \MSUN\ of 
\HI, with a large velocity width, 
as would be expected if the gas is associated with the extended edge-on 
component.  
NED lists three companions within 10\AMIN\ and a further eight, 
many of which are dwarf ellipticals, within 20\AMIN.  
Two of these galaxies are spirals with radial velocities at the edge of
the observed linewidth, and thus might have contributed to the total HI
flux seen by the \GB140. 

\item[C-44~=~NGC~5103~=~UGC~8388] 
We obtain a firm detection of \MHI\ = \hbox{$5.1 \times 10^8$ \MSUN}  
for this system, which optically resembles the Helix galaxy, NGC~2685. 
NED lists no companions within 10\AMIN, and only two optical galaxies
within 20\AMIN, both of which appear too faint and distant to be confused
with NGC~5103. 

\item[C-45~=~Cen~A=~NGC~5128~] 
Cen~A is well-known for its warped lane of gas and dust, which exhibits 
complicated dynamics (Graham 1979, Bland \etal\ 1987, Nicholson \etal\ 
1992).  The \HI\ spectrum is complicated by strong absorption features 
associated with the Milky Way and with Cen~A itself, seen against its strong 
central continuum.  Accordingly, the total \HI\ mass is highly uncertain at 
$7.8\times 10^8$ \MSUN\ plus or minus 30\%, assuming a distance of 3.3 Mpc.  
With an optical size of over 25\AMIN, it is possible that some HI 
flux may lie outside even the \GB140\ beam.
VLA mapping by van~Gorkom \etal\ (1990) showed $3.1 \times 10^8 \msun$ 
of gas, suggesting that there is diffuse or extended emission that was 
missing from the synthesis maps.  
Although owing to its proximity Cen~A is bright in the far-infrared, its 
intrinsic far-infrared luminosity is not remarkable in our sample; furthermore, 
it is relatively gas-poor.

\item[C-46~=~ESO~576-G69] 
This galaxy is wrapped by a series of optical arcs and also exhibits a  
spectacular long tidal tail; the \GB140\ gives 
a 4-$\sigma$ detection  
of $3.7 \times 10^9$ \MSUN\ in \HI\ at a radial velocity 400 \KMS\ larger 
than the optical redshift.    
NED lists three other galaxies of unknown redshifts within 10\AMIN\ of 
ESO~576-G69, as well as an absorption-line system at 5400 \KMS\ seen against 
a background QSO that lies to the southeast of the galaxy behind an optical 
filament. 
Another four galaxies are listed between 10\AMIN\ and 20\AMIN, including 
one ESO~576-G64 with a radial velocity of 5546 \KMS\ that makes it a 
possible source of confusion for the \GB140. 

\item[C-48~=~ESO~326-IG6] 
The \GB140\ gives only a weak upper limit of 
\hbox{$2.1 \times 10^{10}$ \MSUN} of 
\HI\ for this small, relatively distant elliptical with outer 
shell-like optical debris.  NED lists no optical companions with 10\AMIN, 
but three galaxies within 20\AMIN, one of which has 
a radial velocity of 9633 \KMS, placing it just at the edge of our bandwidth.

\item[C-50~=~UGC~10205] 
In many respects, this galaxy looks like a normal Sa galaxy, but its 
edge-on absorption `disk' shows evidence for a brightness bump 
in the light profile (PRC); deep exposures also indicate diffuse debris and 
shell structure in the outer regions (Rubin, 1987).  
The \GB140\ sees $3.9 \times 10^9$ \MSUN\ of \HI, which gives it 
a gas-to-light $\rm M_{\rm HI}/L_{B} = 0.16$.
NED lists no optical companions within the \GB140\ beam.

\item[C-51~=~NGC~6285+6286] 
There is a clear tidal interaction between this close pair; both galaxies 
show optical 
debris that is highly inclined to the their central planes, but 
the more southerly NGC~6286 is the more obvious polar-ring candidate.  
We centered our band on the wrong velocity at the \GB140; the redshift 
listed in PRC is incorrect.  Our bandwidth was large enough, 
however, to cover the optical velocity range of the pair, which allows us 
to place an upper limit of \MHI\ = $4.2 \times 10^9$ \MSUN\ for the system. 
NED lists one companion with 10\AMIN, UGC~10641, which was a radial 
velocity of 5314 \KMS\ and is outside our bandwidth.  In addition, NED 
lists four more galaxies of unknown redshifts within 20\AMIN.  
The pair is infrared luminous; the FIR emission 
is believed to be associated with NGC~6286 (M.~Schmitz, private communication).

\item[C-60~=~ESO~464-G31] 
This disturbed system appears to consist of two nearly perpendicular 
edge-on components, one of which is associated with faint, extended debris.  
GB $140'$ sees a large amount of gas, $9.8 \times 10^9$ \MSUN, with a large 
linewidth, in this somewhat marginal detection.  The automated photometry of 
the ESO-LV catalog (Lauberts \& Valentijn 1989) divided this system into two 
objects of magnitudes 15.02 and 15.64.  We summed these magnitudes 
to obtain the value given in Table 1 and list the larger of the two diameters, 
although it is also possible that this system is a superposition of two 
interacting galaxies on the sky.  
To our knowledge, this represents the first measured redshift of this 
system.  NED lists no other galaxies within the \GB140\ beam.  

\item[C-64~=~ESO~343-IG13] 
The X-shaped optical emission in this galaxy,
with streamers connecting pairs of legs of the X, is believed to 
represent the superposition of two interacting galaxies on the sky.  
The \GB140\ detects $6.2 \times 10^9$ \MSUN\ of \HI\ in a double-horned 
profile.  The system is one of 
the more far-infrared luminous galaxies in our sample.  NED lists one 
other galaxy within 10\AMIN\ and another within 20\AMIN; both are about 
16th magnitude with no listed redshift.   

\item[C-69~=~NGC~7468=UGC~12329] 
The \GB140\ makes a strong detection of $3.6 \times 10^9$ \MSUN\ of \HI\ in 
this system.  NED lists no other companions within 20\AMIN, so this system 
appears to be quite gas-rich, with an $\rm M_{\rm HI}/L_{B} = 0.98$.

\item[D-02~=~NGC~235]
This galaxy, which is sometimes listed as having multiple components, 
is a member of an interacting pair; optical debris 
connects it to its southern partner, NGC~232, located about 2\AMIN\ 
away at a nearly identical redshift.  (In both 
the ESO-LV catalog and NED, two separate components,  
NGC~235A and NGC~235B, are listed with a separation substantially less 
than the optical diameter of the galaxy.  
Checking coordinates against an optical photograph confirms  
that the companion seen in the PRC is NGC~232.)    
NED lists two other galaxies with unknown redshifts within the \GB140\ beam.  
The total \HI\ mass for the system reported by the \GB140\ is 
$5.5 \times 10^9$ \MSUN, some of which may be associated with the neighboring 
galaxies.  Although no far-infrared flux is reported for NGC~235 
by NED, the companion NGC~232 is associated with IRAS flux. 

\item[D-03~=~ESO~474-IG28]
The detection by the \GB140\ of $4.9 \times 10^9$ \MSUN\ of \HI\ is marginal.  
NED lists no companions within 10\AMIN, but three within 20\AMIN:  two with 
discrepant redshifts, and one with unknown redshift.  
The OFF-beam detection at 3059 \KMS\ can be seen as the negative feature 
in the subtracted spectrum.  

\item[D-04~=~ESO~296-G11]
This system, also called the `Boomerang,' 
is thought to be a chance superposition of two galaxies 
on the sky; unlike some of the other objects in our sample, there is no 
evidence for optical tails connecting the two crossed components.  
The ZCAT and RC3 catalogs give optical redshifts of 
5052 \KMS\ and 5572 \KMS\ respectively for this system; interestingly, 
we appear to have detected emission at both 5162 \KMS\ and 5593 \KMS , with 
respective gas masses of $1.5 \times 10^{10}$ \MSUN\ in a broad, asymmetric 
profile and $1.8 \times 10^{9}$ \MSUN\ in a marginal detection the 
magnitude of which is influenced by the choice of baseline.  
NED lists no other optical companions with 10\AMIN, and two galaxies 
within 20\AMIN: one very faint galaxy of unknown redshift, and another 
galaxy with a brightness comparable to ESO~296-G11 with a 
radial velocity of 6576 \KMS.  In order to compute the luminosities and 
their ratios given in two separate lines in Table 3 for this system, 
we have assumed as limiting cases that all of the blue and far-infrared 
flux is associated entirely with one of the two \HI\ redshifts systems.
At least one of these systems must be quite gas-rich.

\item[D-22~=~NGC~4643]
A very faint, disk-like feature is aligned with 
the major axis and extending to three times the optical diameter 
of this otherwise normal-looking barred spiral (\cf PRC).  
The \GB140\ detection 
is at a radial velocity more than 300 \KMS\ lower than the quoted 
optical velocity.  
Since NED lists no optical companions within 10\AMIN, and one small galaxy 
of unknown redshift within 20\AMIN, most of the 
$1.1 \times 10^{8}$ \MSUN\ of \HI\ seen by the \GB140\ is probably 
associated with this system.  Nevertheless, its gas-to-light ratio of
about $\rm M_{\rm HI}/L_{B} = 0.02$ is smaller than average for an
SB0/a (Giovanelli \& Haynes 1988). 

\item[D-25~=~UGC~8387~=~IC~883]
The disturbed morphology of this galaxy includes two linear, one-sided 
and almost perpendicular optical features protruding from center. 
The galaxy is quite infrared-luminous.
The \GB140\ finds $6.1 \times 10^{9}$ \MSUN\ of 
neutral hydrogen.  NED lists one faint galaxy of unknown redshift within 
10\AMIN, and one faint galaxy and a faint galaxy cluster within 20\AMIN.  
Thus, most of the gas is probably associated with UGC~8387 itself.  
Mirabel and Sanders (1988) 
observed this galaxy with the smaller beam of the Arecibo radio telescope 
and found a broad, triple-peaked absorption feature against the 
continuum.  This complicates the interpretation of our emission feature; 
the system may contain more gas than we have estimated.

\item[D-28~=~NGC~6240=~UGC~10592]
This superluminous IRAS galaxy is a merger product: short exposures 
reveal many crossed optical loops and deeper exposures show an extensive 
set of shells.  
The \GB140\ reports a gas mass of $8.2 \times 10^{9}$ \MSUN\ 
within an irregular profile of large linewidth (about 600 \KMS).   
Since this galaxy is a radio source, it is possible that our emission 
spectrum is being influenced by absorption against the continuum.  NED 
lists no optical companions within 10\AMIN, and only one fairly faint 
galaxy within the \GB140\ beam.

\item[D-30~=~ESO~341-IG4]
The \GB140\ reports a firm detection of $1.4 \times 10^{10}$ \MSUN\ of 
\HI\ in this system, which according to NED has no neighbors within 10\AMIN.   
One galaxy and a galaxy cluster, both at higher redshift, do lie within the 
\GB140\ beam.  

\item[D-35~=~NGC~7252]
This, the famous `Atoms for Peace' galaxy studied by Schweizer (1982), 
is the most optically-luminous and one of the more infrared-luminous 
galaxies in our sample.  The galaxy is an obvious merger product, as 
evidenced by the remarkable optical loops and tidal tails surrounding it.  
The \GB140\ detects $4.5 \times 10^{9}$ \MSUN\ of \HI, which given 
the large blue luminosity of this galaxy, makes it gas-poor compared 
to others in our sample.  The linewidth of 200 \KMS\ seems  
surprisingly low given the disturbed optical morphology of the system.  
VLA observations (Hibbard \etal\ 1993) show 
$3.6 \times 10^{9}$ \MSUN\ of \HI\ in NGC~7252, tails and all, 
and a further $1.4 \times 10^{9}$ \MSUN\ in an uncatalogued companion, 
A2218-248, about 12\AMIN\ and 50 \KMS\ away.  
The sum of these two fluxes is within 15\% of our measured value.

\item[D-43~=~ESO~510-G13]
This boxy system has a very thin optical disk and an associated dust 
lane along its major axis.  Probably for this reason, the 
automated photometry routines of ESO-LV catalog have broken the system into two
fainter `galaxies,' one associated with the flat disk and one with the
boxy feature.  We have therefore used the magnitude and diameter quoted
from RC3 for inclusion in Table~1.  The \GB140\ reports an 
upper limit for the \HI\ mass of this system of $3.3 \times 10^{9}$ \MSUN.  
No other galaxies are reported by NED to be within the \GB140\ beam.

\end{description}

\vfill\eject
\end{document}


\pagestyle{empty}
\samepage
\font\ninept=cmr9
\hsize=10in
\vsize=6.5in
\hoffset-1.25in
\voffset-1.45in
\newcommand{\etal}{{\it et~al.}}                
\newcommand{\KMS}{$\rm km\ s^{-1}$}
%
%
%
%
\begin{tabular}{|c|l|ll|rr|lr|lr|r|rl|c|rr|}
\multicolumn{16}{c}{{\bf Table~1}.\ 
Basic~Optical~and~HI~Data~for~the~Green~Bank~Sample~of~Polar-Ring~Galaxies}\\ [4pt]\hline
&&&&&&&&&&&&&&&\\
\noalign{\vskip-6pt}
PRC&\multicolumn{1}{c|}{Name}&\multicolumn{1}{c}{R.A.}
&\multicolumn{1}{c|}{Dec.}&\multicolumn{2}{c|}{v$_{\hbox{\ninept opt}}$}
&\multicolumn{2}{c|}{B$_{\hbox{\ninept T}}$}&\multicolumn{2}{c|}{Log}
&\multicolumn{1}{c|}{v$_{\rm HI}$}&\multicolumn{2}{c|}{$\int S * dv$}
&\multicolumn{1}{c|}{$\Delta {\rm v_{20}}$}
&\multicolumn{1}{c}{S$_{\hbox{\ninept max}}$}
&\multicolumn{1}{c|}{S$_{\hbox{\ninept rms}}$}
\\
No.&&\multicolumn{2}{c|}{(1950.0)}
&\multicolumn{2}{c|}{[km s$^{-1}$]}
&\multicolumn{2}{c|}{[$^m$]}
&\multicolumn{2}{c|}{\hfil D$_{25} / 0.1'$\hfil}
&\multicolumn{1}{c|}{[km s$^{-1}$]}
&\multicolumn{2}{c|}{[Jy $\cdot$ km s$^{-1}$]}
&\multicolumn{1}{c|}{[km s$^{-1}$]}
&\multicolumn{2}{c|}{[Jy]}\\
(1)&\multicolumn{1}{c|}{(2)}&\multicolumn{1}{c}{(3)}
&\multicolumn{1}{c|}{(4)}&\multicolumn{1}{c}{(5)}&\multicolumn{1}{c|}{(6)}
&\multicolumn{1}{c}{(7)}&\multicolumn{1}{c|}{(8)}
&\multicolumn{1}{c}{(9)}&\multicolumn{1}{c|}{(10)}
&\multicolumn{1}{c|}{(11)}&\multicolumn{1}{c}{(12)}&\multicolumn{1}{c|}{(13)}
&\multicolumn{1}{c|}{(14)}&\multicolumn{1}{c}{(15)}&\multicolumn{1}{c|}{(16)}
\\ [4pt]\hline
&&&&&&&&&&&&&&&\\
\noalign{\vskip-6pt}
A-01&A0136$$-$$0801&01\ 36\ 25.4&$-$08\ 01\ 24
&5528&S&15.82&P&1.0&T&5521&3.82&
1.2&408&0.018&0.0029\\
A-02&ESO~415$-$G26&02\ 26\ 11.5&$-$32\ 06\ 14
&4560&W&14.7&R&1.12&R&4572&
6.22&1.8&441&0.029&0.0040\\
A-03&NGC~2685&08\ 51\ 41.2&+58\ 55\ 30
&869&R&12.12&R&1.65&R&875&34.2$\phantom{0}$&4.0&298&0.156&0.0145\\
&NGC~2685~OFF$^{\ninept 1}$&08\ 41\ 11&+58\ 55\ 30
&(1288)&R&13.98&R&&&1288&6.87&0.87&
 151&0.069&0.0091\\
A-05&NGC~4650A&12\ 42\ 04.8&$-$40\ 26\ 35
&2904&W&14.27&P&1.20&R&2909&23.2$\phantom{0}$&2.4
&244&0.138&0.0098\\
A-06&UGC~9796&15\ 14\ 00&+43\ 22\ 00
&5606&Z&16.04&P&1.14&R&5423&16.4$\phantom{0}$&1.7
&354&0.084&0.0047\\ [6pt]\hline 
&&&&&&&&&&&&&&&\\
\noalign{\vskip-4pt}
B-01&IC~$51^{\ninept 2}$&00\ 43\ 53&$-$13\ 42\ 58
&1758&H&13.&N&1.13&R&1709&11.5$\phantom{0}$&1.2&253
&0.073&0.0047 \\
B-11&UGC~5600&10\ 19\ 16.6&+78\ 52\ 51
&2823&R&14.19&R&1.15&R&2769&12.0$\phantom{0}$&1.3
&173&0.098&0.0076\\
B-12&ESO~503$-$G17&11\ 24\ 24&$-$27\ 25\ 48
&10313&J&16.54&P&0.85&LV&9800$-$11900&
$<5.48$&&&$<0.022$&0.0073\\
B-16&NGC~5122$^{\ninept 3}$&13\ 21\ 36&$-$10\ 23\ 39
&2861&J& & & & &
4804&3.96$\phantom{0}$&1.2&293&0.022&0.0040\\   
[8pt]
B-17&UGC~9562&14\ 49\ 13.1&+35\ 44\ 53
&1236&R&14.41&R&0.98&R&1239&13.6$\phantom{0}$&1.6
&200&0.105&0.0080  \\ 
B-21&ESO~603$-$G21&22\ 48\ 41&$-$20\ 30\ 42
&3150&P&15.5&LV&0.92&LV&3180&14.3$\phantom{0}$&1.3
&272&0.069&0.0047\\
B-23&A2330$-$3751&23\ 30\ 01&$-$37\ 51\ 09
&&&&&1.1&T&1000$-$8000&
$<2.18$&&&$<0.009$&0.0029\\
B-27&ESO~293$-$IG17$^{\ninept 4}$&23\ 53\ 53&$-$39\ 26\ 42
&&&16.17&LV&0.86&LV&1000$-$8000&
$<2.45$&&&$<0.010$&0.0033\\ 
[6pt] \hline
&&&&&&&&&&&&&&&\\
\noalign{\vskip-4pt}
C-09&NGC~442&01\ 12\ 05.1&$-$01\ 17\ 05
&5620&R&14.45&R&0.99&R&5629&1.53&0.9
&302&0.011&0.0029\\
C-11&NGC~625&01\ 32\ 55.0&$-$41\ 41\ 18
&383&R&11.71&R&1.76&R
&394&37.4$\phantom{0}$&4.3&113&0.513&0.0167\\
C-12&UGC~1198&01\ 40\ 58&+85\ 00\ 38
&1207&H&14.8&R&1.00&R&1149&1.64&1.3
&174&0.033&0.0073\\
C-13&NGC~660&01\ 40\ 20.8&+13\ 23\ 20
&823&R&12.02&R&1.92&R&848&187.8$\phantom{0}$&\llap{5}0&325&0.909&0.1553\\
[8pt]
C-14&NGC~979&02\ 29\ 46.2&$-$44\ 44\ 38&&
&13.78&R&1.07&R&4775&26.2$\phantom{0}$&7.4
&678&0.055&0.0109  \\ 
C-18&ESO~358$-$G20&03\ 32\ 53&$-$32\ 47\ 42
&1853&R&14.54&R&1.08&R&1799&1.75&
0.6&167&0.018&0.0036\\
C-24&UGC~4261&08\ 07\ 40.3&+36\ 58\ 40
&6419&H&14.78&R&0.96&R&6410&4.62&0.8
&241&0.029&0.0033\\
C-25&UGC~4323&08\ 15\ 36.6&+67\ 08\ 22
&3981&H&14.08&R&1.20&R&3686&3.56&1.0
&397&0.018&0.0025\\
&UGC~4323~OFF&&\ &\ &&\ &&\ &&4785&1.35&0.3&116&0.015&0.0025\\
[8pt]
C-26&UGC~4332&08\ 16\ 43.1&+21\ 16\ 22
&5505&R&14.82&R&1.12&R&5811&3.96&1.4
&299&0.022&0.0047\\
&UGC~4332~OFF&08\ 27\ 13&+21\ 16\ 22&&&&&&
&5134&1.67&0.5&104&0.022&0.0044 \\ 
C-27&UGC~4385&08\ 21\ 04.2&+14\ 54\ 56
&1969&Z&14.51&R&0.94&R&1954&7.93&0.9
&196&0.062&0.0047\\
C-29&NGC~2865&09\ 21\ 14.6&$-$22\ 56\ 47
&2611&R&12.57&R&1.39&R&2772&2.95&0.40
&164&0.036&0.0051\\
C-35&NGC~3414&10\ 48\ 31.8&+28\ 14\ 28
&1434&R&11.96&R&1.55&R&400$-$2300&$<7.13$&&&$<0.029$&0.0095\\
[8pt]
C-38&NGC~3934&11\ 49\ 38&+17\ 07\ 44
&3626&R&14.5&R&1.03&R&3793&5.71&0.7
&91&0.087&0.0080\\
C-39&NGC~4174&12\ 09\ 54.9&+29\ 25\ 42
&3980&R&14.31&R&0.92&R&3811&12.9$\phantom{0}$&4.0
&439&0.062&0.0091\\ 
C-41&IC~3370&12\ 24\ 58&$-$39\ 03\ 41
&2930&R&11.99&R&1.46&R&3219&1.42&0.7&92&0.025&0.0076\\
C-42&NGC~4672&12\ 43\ 30&$-$41\ 26\ 00
&3389&R&14.09&R&1.31&R&3242&7.67&2.5&401&0.040&0.0062\\
[4pt]\hline
\end{tabular}
\vfil\eject
\begin{tabular}{|c|l|ll|rr|lr|lr|r|rl|c|rr|}
\multicolumn{16}{c}{{\bf Table~1}.\ (continued)}\\ [4pt]\hline
&&&&&&&&&&&&&&&\\
\noalign{\vskip-6pt}
PRC&\multicolumn{1}{c|}{Name}&\multicolumn{1}{c}{R.A.}
&\multicolumn{1}{c|}{Dec.}&\multicolumn{2}{c|}{v$_{\hbox{\ninept opt}}$}
&\multicolumn{2}{c|}{B$_{\hbox{\ninept T}}$}&\multicolumn{2}{c|}{Log}
&\multicolumn{1}{c|}{v$_{\rm HI}$}&\multicolumn{2}{c|}{$\int S * dv$}
&\multicolumn{1}{c|}{$\Delta {\rm v_{20}}$}
&\multicolumn{1}{c}{S$_{\hbox{\ninept max}}$}
&\multicolumn{1}{c|}{S$_{\hbox{\ninept rms}}$}
\\
No.&&\multicolumn{2}{c|}{(1950.0)}
&\multicolumn{2}{c|}{[km s$^{-1}$]}
&\multicolumn{2}{c|}{[$^m$]}
&\multicolumn{2}{c|}{\hfil D$_{25} / 0.1'$\hfil}
&\multicolumn{1}{c|}{[km s$^{-1}$]}
&\multicolumn{2}{c|}{[Jy $\cdot$ km s$^{-1}$]}
&\multicolumn{1}{c|}{[km s$^{-1}$]}
&\multicolumn{2}{c|}{[Jy]}\\
(1)&\multicolumn{1}{c|}{(2)}&\multicolumn{1}{c}{(3)}
&\multicolumn{1}{c|}{(4)}&\multicolumn{1}{c}{(5)}&\multicolumn{1}{c|}{(6)}
&\multicolumn{1}{c}{(7)}&\multicolumn{1}{c|}{(8)}
&\multicolumn{1}{c}{(9)}&\multicolumn{1}{c|}{(10)}
&\multicolumn{1}{c|}{(11)}&\multicolumn{1}{c}{(12)}&\multicolumn{1}{c|}{(13)}
&\multicolumn{1}{c|}{(14)}&\multicolumn{1}{c}{(15)}&\multicolumn{1}{c|}{(16)}
\\ 
[4pt]\hline
&&&&&&&&&&&&&&&\\
\noalign{\vskip-6pt}
C-44&NGC~5103&13\ 18\ 17.6&+43\ 20\ 45
&1283&R&13.6&R&1.16&R&1289&6.76&
0.9&141& 0.062 &0.0062\\
C-45&NGC~5128&13\ 22\ 31.8&$-$42\ 45\ 30
&538&R&\phantom{1}7.84&R&2.41&R
&524&300$\phantom{.00}$&\llap{10}0&570&0.822&0.0894\\
C-46&ESO~576$-$G69&13\ 27\ 22&$-$20\ 40\ 36
&5336&H&14.6&LV&0.95&LV&5759&2.91&0.9
&109&0.033&0.0084\\
C-48&ESO~326$-$IG6&14\ 08\ 04.7&$-$39\ 52\ 15
&8584&R&15.&E&1.23&R&7600$-$9600&$<7.13$
&&&$<0.029$&0.0095\\ 
[8pt]
C-50&UGC~10205&16\ 04\ 40.6&+30\ 14\ 06
&6605&R&14.4&R&1.16&R&6773&1.96&0.5
&163&0.015&0.0029\\
C-51&NGC~6285+6286$^{\ninept 5}$&16\ 57\ 47&+58\ 59\ 47
&5595&R&14.06&R&1.12&R&5350$-$7350&
$<3.00$&&&$<0.012$&0.0040\\
&NGC~6285/6~OFF&&\ &\ &&\ &&\ &
&5639&2.47&0.5&121&0.029&0.0040\\
C-60&ESO~464$-$G31$^{\ninept 6}$&21\ 15\ 25&$-$27\ 33\ 35
&&&14.5&LV&1.07&LV&6529&5.42&1.9
&573&0.011&0.0033\\
C-64&ESO~343$-$IG13$^{\ninept 7}$&21\ 33\ 05&$-$38\ 46\ 06
&5685&R&14.55&R&1.17&R&5716&4.55&1.5
&342&0.022&0.0044\\
&&&&5822&R&&&&&&&&&&\\
C-69&NGC~7468&23\ 00\ 30.3&+16\ 20\ 04
&2123&R&14.16&R&0.96&R&2072&16.1$\phantom{0}$&1.2
&247&0.109&0.0047\\ 
[6pt] \hline
&&&&&&&&&&&&&&&\\
\noalign{\vskip-4pt}
D-02&NGC~235
&00\ 40\ 24&$-$23\ 48\ 55
&6664&R&14.08&R&1.12&R&6710&2.84&1.1
&282&0.015&0.0040\\
D-03&ESO~474$-$IG28&00\ 45\ 06&$-$23\ 04\ 24&6780&H
&14.91&LV&0.87&LV&6610&2.62&1.0
&220&0.025&0.0044\\
&ESO~474$-$IG28 OFF&&\ &&\ &&\ &\ &
&3059&5.31&1.6&360&0.033&0.0044\\
D-04&ESO~296$-$G11$^{\ninept 8}$&01\ 17\ 43&$-$41\ 29\ 55
&5052&Z&14.5&R&1.01&R&5162&
14.0$\phantom{0}$&2.6&442&0.051&0.0058\\
&&&&5572&R&&&&&5593&1.38&0.9&167&0.018&0.0051\\
D-22&NGC~4643&12\ 40\ 46.9&+02\ 15\ 06
&1399&R&11.72&R&1.49&R&1053&3.20&0.8&141&0.029&0.0055\\
&NGC~4643~OFF&12\ 30\ 22\ &+02\ 15\ 00\ &&&&&&
&2199&1.67&0.7&
136&0.029&0.0051\\ 
[8pt]
D-25&UGC~8387&13\ 18\ 17&+34\ 24\ 03
&6892&R&14.4&R&1.19&R&7024&2.95&1.0
&279&0.018&0.0036\\
D-28&NGC~6240&16\ 50\ 27.7&+02\ 29\ 00
&7351&R&13.8&R&1.33&R&7197&3.78&1.4&616&0.018&0.0022\\
D-30&ESO~341$-$IG4&20\ 37\ 59&$-$38\ 22\ 20
&5973&R&13.38&R&1.28&R&6055&9.05&1.3
&284&0.051&0.0047\\
D-35&NGC~7252&22\ 17\ 57.7&$-$24\ 55\ 51
&4760&R&12.59&R&1.29
&R&4752&4.58&0.56&191&0.040&0.0044\\
D-43&ESO~510$-$G13&13\ 52\ 14&$-$26\ 32\ 18
&3559&R&13.35&R&1.29&R&2600$-$4500&$<7.13$
&&&$<0.029$&0.0095\\ 
[4pt] \hline
\noalign{\vskip8pt}
\noalign{\vbox{\hsize=9.5in\noindent\baselineskip=12pt

The letters listed in Columns 6, 8 and 10 of Table 1 refer to the 
following sources for optical redshifts, blue magnitudes and major 
axis diameters:  
S = SWR; 
W = WMS; 
R = RC3; 
Z = ZCAT, 1992; 
H = J.~Huchra, private communication 1993; 
J = Jarvis \& Sackett, unpublished; 
P = PRC; 
N = NBG; 
RN = Revised NGC;
LV = ESO-LV Catalog, Lauberts \& Valentijn 1989; 
E = ESO Catalog, Lauberts, 1982; and 
T = This work; estimates of optical size made from photographs 
published in the PRC.
\smallskip
\hfil\break

$^{\ninept 1}$\ We identify the signal in the off-beam of NGC~2685 as  
	UGC~4549.  Independent {\it radio\/} observations give the 
	same velocity; no optical redshift was available.  \hfil\break
$^{\ninept 2}$\ The coordinates quoted in the PRC for IC~51 are incorrect; 
      we give the correct coordinates here.\hfil\break
$^{\ninept 3}$\ Due to an unreliable redshift, our Green Bank bandpass 
	missed this galaxy, but did make an unrelated detection at 
	about 4800 \KMS.  
	\hfil\break
$^{\ninept 4}$\ Flux blending from neighbors make the total B magnitude 
	reported from the ESO-LV catalog for this system somewhat uncertain.  
	\hfil\break
$^{\ninept 5}$\ NGC~6285/6286 is an interacting pair; NGC~6286 is the 
	polar-ring candidate.  NGC~6285 has a heliocentric velocity of 
	5696 \KMS\ and $m_{\rm B} = 14.48$.   \hfil\break
$^{\ninept 6}$\ Automated photometry broke this system into two objects; 
	see notes for further explanation.\hfil\break
$^{\ninept 7}$\ ESO~343-IG13 is an interacting pair of galaxies that are 
	superimposed on the sky.\hfil\break
$^{\ninept 8}$\ ESO~296-G11 is a superposition of two galaxies.  We have 
      found two optical redshifts in the literature; apparently our survey 
      detected both galaxies in HI.\hfil\break
}}
\end{tabular}
\clearpage


\pagestyle{empty}
\samepage
\font\ninept=cmr9
\hsize=10in
\vsize=6.5in
\hoffset-1.25in
\voffset-1.25in
\def\ie{{\it ie.,\/}}
%
\begin{tabular}{|c|l|c|rrrr|cccc|c|ccc|}
\multicolumn{15}{c}{{\bf Table~2}.\ 
Basic~Infrared~Data~for~the~Green~Bank~Sample~of~Polar-Ring~Galaxies}\\ [4pt]\hline
&&&&&&&&&&&&&&\\
\noalign{\vskip-6pt}
PRC
&\multicolumn{1}{c|}{Galaxy}
&\multicolumn{1}{c|}{IRAS}
&\multicolumn{4}{c|}{Flux Densities~[Jy]}
&\multicolumn{4}{c|}{Flux Qualities}
&\multicolumn{1}{c|}{Log(F$_{\rm {\ninept FIR}}$)}
&\multicolumn{3}{c|}{Infrared Colors}\\
No.
&\multicolumn{1}{c|}{Name}
&\multicolumn{1}{c|}{Reference}
&\multicolumn{1}{c}{12$\mu$}
&\multicolumn{1}{c}{25$\mu$}
&\multicolumn{1}{c}{60$\mu$}
&\multicolumn{1}{c|}{100$\mu$}
&\multicolumn{1}{c}{12$\mu$}
&\multicolumn{1}{c}{25$\mu$}
&\multicolumn{1}{c}{60$\mu$}
&\multicolumn{1}{c|}{100$\mu$}
&\multicolumn{1}{c|}{[W m$^{-2}$]}
&\multicolumn{1}{c}{(12/25)}
&\multicolumn{1}{c}{(25/60)}
&\multicolumn{1}{c|}{(60/100)}
\\
(1)
&\multicolumn{1}{c|}{(2)}
&\multicolumn{1}{c|}{(3)}
&\multicolumn{1}{c}{(4)}
&\multicolumn{1}{c}{(5)}
&\multicolumn{1}{c}{(6)}
&\multicolumn{1}{c|}{(7)}
&\multicolumn{1}{c}{(8)}
&\multicolumn{1}{c}{(9)}
&\multicolumn{1}{c}{(10)}
&\multicolumn{1}{c|}{(11)}
&\multicolumn{1}{c|}{(12)}
&\multicolumn{1}{c}{(13)}
&\multicolumn{1}{c}{(14)}
&\multicolumn{1}{c|}{(15)}\\ 
[4pt]\hline
&&&&&&&&&&&&&&\\
\noalign{\vskip-6pt}

A-01 & A 0136-0801   &	    	&	  &		&	 &	  &	&	&	&   &	  &    &          &	    \\
A-02 & ESO 415-G26   &   	&	  &		&	 &	  &	&	&	&   &	  &    &          &	    \\
A-03 & NGC 2685      & FSC2	&$\phantom{00}$0.09842 & $<0.0975\rlap{8}$	&$\phantom{0}$0.3156 & 1.694$\phantom{0}$  & 3	& 1	& 3	& 2 & $-$13.50 &  &	  & $-$0.730\\
A-05 & NGC 4650A     & FSC2	&$\llap{$<$}0.09979$& $<0.1515$ & 0.3662 &$<1.503\phantom{0}$& 1	& 1	& 3	& 1 & $\llap{$<$}-13.51$ 
&  &	  &	    \\
A-06 & UGC 9796$^{\ninept 1}$&		&	  &		&	 &	  &    	&	&	&   &	   &    &          &	    \\
[6pt]\hline 
&&&&&&&&&&&&&&\\
\noalign{\vskip-4pt}
B-01 & IC 51	     & FSC2	&0.1557$\phantom{0}$  & 0.2004 	& 2.214$\phantom{0}$  & 4.691$\phantom{0}$ & 2	& 2	& 3	& 2 & $-$12.88 & $-$0.110 & $-$1.043 & $-$0.326\\
B-11 & UGC 5600      & FSC2	&0.1544$\phantom{0}$  & 0.6052	& 3.464$\phantom{0}$  & 4.961$\phantom{0}$   & 3	& 3	& 3	& 2 & $-$12.76 & $-$0.593 & $-$0.758 & $-$0.156\\
B-12 & ESO 503-G17   &	   	& 	  &		&	 &	  &   	&	&	&   &	   &   &          &	    \\
[8pt]
B-17 & UGC 9562      &	  	& 	  &		&	 &	  &   	&	&	&   &	   &   &          &	    \\
B-21 & ESO 603-G21   & FSC2	&0.1574$\phantom{0}$  &$<0.1530$	& 1.454$\phantom{0}$  & 2.889$\phantom{0}$  & 2	& 1	& 3	& 2 & $-$13.08 &  &	  & $-$0.298\\
B-23 & A 2330-3751$^{\ninept 2}$&  	& 	  &		&	 &	  &   	&	&	&   &	    &  &          &	    \\
B-27 & ESO 293-IG17  &	  	& 	  &		&	 &	  &   	&	&	&   &	    &  &          &	    \\
[6pt]\hline 
&&&&&&&&&&&&&&\\ 
\noalign{\vskip-4pt}												
C-09 & NGC  442$^{\ninept 2}$&	  	& 	  &		&	 &	  &   	&	&	&   &	    &  &          &	    \\
C-11 & NGC 625       & FSC2	&0.1997$\phantom{0}$  & 0.9473      & 5.094$\phantom{0}$  & 9.079$\phantom{0}$  & 3	& 3	& 3	& 2 & $-$12.55 & $-$0.676 & $-$0.731 & $-$0.251\\
C-12 & UGC 1198      & FSC2	&0.06611 & 0.4148      & 2.53$\phantom{00}$& 2.958$\phantom{0}$  & 2	& 3	& 3	& 2 & $-$12.92 & $-$0.798 & $-$0.785 & $-$0.068\\
C-13 & NGC 660       & FSC2	&2.421$\phantom{00}$& 7.526$\phantom{0}$ & 65.54$\phantom{00}$ & 102.04$\phantom{00}$ & 2	& 3	& 3	& 2 & $-$11.47 & $-$0.493 & $-$0.940 & $-$0.192\\
     & 		     & SCANS	&2.88$\phantom{000}$& 7.71$\phantom{00}$& 69.92$\phantom{00}$  & 101.5$\phantom{000}$  & R &	MR &	MR	& UR& $-$11.45 & $-$0.428 & $-$0.958 & $-$0.162\\
[8pt]
C-14 & NGC 979       &	    	& 	  &		&	 &	  &   	&	&	&   &	   &   &          &	    \\
C-18 & ESO 358-G20   &	    	& 	  &		&	 &	  &   	&	&	&   &	   &   &          &	    \\
C-24 & UGC 4261      & FSC2	&$\llap{$<$}0.1015\phantom{0}$& 0.1553      & 1.143$\phantom{0}$ & 1.345$\phantom{0}$  & 1	& 2	& 3	& 2 & $-$13.27 &  & $-$0.867 & $-$0.071\\
C-25 & UGC 4323$^{\ninept 2}$&	    	& 	  &		&	 &	  &   	&	&	&   &	   &  &          &	    \\
[8pt]
C-26 & UGC 4332      & FSC2	&$\llap{$<$}0.1995\phantom{0}$& 0.2315	& 0.8074 & 2.089$\phantom{0}$  & 1	& 2	& 3	& 2 & $-$13.28 &  & $-$0.543 & $-$0.413\\
C-27 & UGC 4385      & FSC2	&$\llap{$<$}0.07956$& $<0.2213$ & 0.3124 & 0.8685 & 1	& 1	& 3	& 2 & $-$13.68 &   &          & $-$0.444\\
C-29 & NGC 2865      &	   	& 	  &		&	 &	  &   	&	&	&   &	  &    &          &	    \\
C-35 & NGC 3414      & FSC2	&$\llap{$<$}0.1161\phantom{0}$& $<0.1556$  & 0.1907 &$<0.9129$& 1	& 1	& 3	& 1 & $\llap{$<$}-13.75$ 
&  &          &	    \\
[8pt]
C-38 & NGC 3934      & FSC2	&0.1242$\phantom{0}$& $<0.1553$   & 2.22$\phantom{00}$ & 4.277$\phantom{0}$  & 2	& 1	& 3	& 2 & $-$12.90 &  &          & $-$0.285\\
C-39 & NGC 4174$^{\ninept 3}$&	    	& 	  &		&	 &	  &   	&	&	&   &	      &  &          &         \\
C-41 & IC 3370       & FSC2	&$\llap{$<$}0.1244\phantom{0}$& $<0.1010$  & 0.5559 & 1.595$\phantom{0}$  & 1	& 1	& 3	& 2 & $-$13.42 &  &          & $-$0.458\\
C-42 & NGC 4672      & FSC2	&0.1452$\phantom{0}$  & 0.2263	& 2.391$\phantom{0}$ & 6.641$\phantom{0}$  & 3	& 3	& 3	& 2 & $-$12.79 & $-$0.193 & $-$1.024 & $-$0.444\\
[4pt]\hline
\end{tabular}
\vfil\eject
\begin{tabular}{|c|l|c|rrrr|cccc|c|ccc|}
\multicolumn{14}{c}{{\bf Table~2}.\ (continued)}\\ [4pt]\hline
&&&&&&&&&&&&&&\\
\noalign{\vskip-6pt}
PRC
&\multicolumn{1}{c|}{Galaxy}
&\multicolumn{1}{c|}{IRAS}
&\multicolumn{4}{c|}{Flux Densities~[Jy]}
&\multicolumn{4}{c|}{Flux Qualities}
&\multicolumn{1}{c|}{Log(F$_{\rm {\ninept FIR}}$)}
&\multicolumn{3}{c|}{Infrared Colors}\\
No.
&\multicolumn{1}{c|}{Name}
&\multicolumn{1}{c|}{Reference}
&\multicolumn{1}{c}{12$\mu$}
&\multicolumn{1}{c}{25$\mu$}
&\multicolumn{1}{c}{60$\mu$}
&\multicolumn{1}{c|}{100$\mu$}
&\multicolumn{1}{c}{12$\mu$}
&\multicolumn{1}{c}{25$\mu$}
&\multicolumn{1}{c}{60$\mu$}
&\multicolumn{1}{c|}{100$\mu$}
&\multicolumn{1}{c|}{[W m$^{-2}$]}
&\multicolumn{1}{c}{(12/25)}
&\multicolumn{1}{c}{(25/60)}
&\multicolumn{1}{c|}{(60/100)}
\\
(1)
&\multicolumn{1}{c|}{(2)}
&\multicolumn{1}{c|}{(3)}
&\multicolumn{1}{c}{(4)}
&\multicolumn{1}{c}{(5)}
&\multicolumn{1}{c}{(6)}
&\multicolumn{1}{c|}{(7)}
&\multicolumn{1}{c}{(8)}
&\multicolumn{1}{c}{(9)}
&\multicolumn{1}{c}{(10)}
&\multicolumn{1}{c|}{(11)}
&\multicolumn{1}{c|}{(12)}
&\multicolumn{1}{c}{(13)}
&\multicolumn{1}{c}{(14)}
&\multicolumn{1}{c|}{(15)}\\ 
[4pt]\hline
&&&&&&&&&&&&&&\\
\noalign{\vskip-6pt}
C-44 & NGC 5103      &	    	& 	  &		&	 &        &	&	&   	&   &      &   &	  &	    \\
C-45 & NGC 5128      & FSC2	&$\phantom{0}$13.26$\phantom{000}$& 17.26$\phantom{00}$& \llap{16}2.2$\phantom{000}$	 & 313.8$\phantom{000}$  & 3	& 3	& 2	& 2 & $-$11.03 & $-$0.114 & $-$0.973 & $-$0.287\\
     & 	             & MAP	&23.03$\phantom{000}$&30.74$\phantom{00}$& \llap{21}7.6$\phantom{000}$  &501.2$\phantom{000}$  &  	&  	&  	&   & $-$10.87 & $-$0.125 & $-$0.850 & $-$0.362\\
C-46 & ESO 576-G69   & FSC2	&$\llap{$<$}0.1192\phantom{0}$& 0.1568	& 0.3609 & 1.581$\phantom{0}$  & 1	& 2	& 3	& 2 & $-$13.50 &  & $-$0.362 & $-$0.642\\
C-48 & ESO 326-IG6   &	   	& 	  &		&	 &	  &   	&	&	&   &	  &    &          &	    \\
[8pt]
C-50 & UGC 10205     & FSC2	&$\llap{$<$}0.08905$&$<0.0841\rlap{1}$ & 0.3939 & 1.538$\phantom{0}$  & 1	& 1	& 3	& 2 & $-$13.49 &  &          & $-$0.592\\
C-51 & NGC 6285+6286$^{\ninept 4}$& FSC2	&0.3305$\phantom{0}$  & 0.4941	& 7.8778 & \llap{2}2.60$\phantom{00}$ & 3	& 3	& 3	& 2 & $-$12.27 & $-$0.175 & $-$1.203 & $-$0.458\\
     & 	             & SCANS	&0.50$\phantom{000}$& 0.64$\phantom{00}$& 9.87$\phantom{00}$   & 22.01$\phantom{00}$  & R	& R	& MR	& UR& $-$12.22 & $-$0.107 & $-$1.188 & $-$0.348\\
C-60 & ESO 464-G31   &	   	& 	  &		&	 &	  &   	&	&	&   &	  &  &            &	    \\
C-64 & ESO 343-IG13  & FSC2	&0.2795$\phantom{0}$  & 0.8804	& 5.855$\phantom{0}$	 & 8.906$\phantom{0}$  & 3	& 3	& 3	& 2 & $-$12.52 & $-$0.498 & $-$0.823 & $-$0.182\\
C-69 & NGC 7468      & FSC2	&$\llap{$<$}0.09844$& $<0.1439$ & 1.250$\phantom{0}$  & 1.489$\phantom{0}$ & 1	& 1	& 3	& 2 & $-$13.23 &  &	  & $-$0.076\\
[6pt]\hline 
&&&&&&&&&&&&&&\\
\noalign{\vskip-4pt}
D-02 & NGC 235$^{\ninept 5}$&	   	& 	  &		&	 &	  &   	&	&	&   &	    &  &          &         \\
D-03 & ESO 474-IG28  & FSC2	&$\llap{$<$}0.09424$&$<0.0815\rlap{8}$	& 0.2417 &$<0.6439$& 1	& 1	& 3	& 1 & $\llap{$<$}-13.80$
&  &	  &         \\
D-04 & ESO 296-G11$^{\ninept 6}$& FSC2	&$\llap{$<$}0.08711$& 0.0933\rlap{8}& 0.8823 & 1.578$\phantom{0}$  & 1	& 2	& 3	& 2 & $-$13.31 &  & $-$0.975 & $-$0.252\\
D-22 & NGC 4643      & FSC2	&$\llap{$<$}0.09545$& $<0.1077$ & 0.5018 & 1.858$\phantom{0}$  & 1	& 1	& 3	& 2 & $-$13.40 &  & 	  & $-$0.569\\
[8pt]
D-25 & UGC 8387      & FSC2	&0.2628$\phantom{0}$  & 1.362$\phantom{0}$& 15.44$\phantom{00}$  & 25.18$\phantom{00}$ & 3	& 2	& 2	& 2 & $-$12.09 & $-$0.714 & $-$1.055 & $-$0.212\\
     & 	             & SCANS	&0.26$\phantom{000}$& 1.37$\phantom{00}$& 33.69$\phantom{00}$  & 24.9$\phantom{000}$  & UR	& UR	& UR	& UR& $-$12.12 & $-$0.722 & $-$1.000 & $-$0.260\\
D-28 & NGC 6240      & FSC2	&0.5578$\phantom{0}$  & 3.417$\phantom{0}$& 22.68$\phantom{00}$  & 7.78$\phantom{00}$	  & 3	& 3	& 3	& 2 & $-$11.96 & $-$0.787 & $-$0.822 & $-$0.088\\
D-30 & ESO 341-IG4   & FSC2	&0.1189$\phantom{0}$  & 0.2719	& 1.308$\phantom{0}$ & 1.858$\phantom{0}$  & 2	& 3	& 3	& 2 & $-$13.18 & $-$0.359 & $-$0.682 & $-$0.152\\
D-35 & NGC 7252      & FSC2	&0.2398$\phantom{0}$ & 0.4322	& 3.975$\phantom{0}$ & 7.017$\phantom{0}$  & 3	& 3	& 3	& 2 & $-$12.66 & $-$0.256 & $-$0.964 & $-$0.247\\
D-43 & ESO 510-G13   & FSC2	&$\llap{$<$}0.1598\phantom{0}$& $<0.2401$  & 1.682$\phantom{0}$  & 6.063$\phantom{0}$  & 1	& 1	& 3	& 2 & $-$12.88 &  & 	  & $-$0.557\\
[6pt]\hline 
\noalign{\vskip8pt}
\noalign{\vbox{\hsize=9.5in\noindent\baselineskip=12pt

The flux qualities listed in columns 8-11 are taken from Version 2 of the IRAS Faint Source Catalog (FSC2), 
where 1 is low quality (\ie upper limit) and 3 is high quality.  For systems in which integrated fluxes 
from scans or IR maps were available, the columns 8-11 indicate whether the emission was resolved with 
IRAS:  
R = resolved;
MR = marginally resolved;
UR = unresolved.  Beam sizes are $0.77'$, $0.78'$, $1.44'$, and $2.94'$ for the 12$\mu$, 25$\mu$, 60$\mu$, 
and 100$\mu$ bands, respectively.
\smallskip
\hfil\break

$^{\ninept 1}$\ Although no flux is reported in any IRAS band for UGC~9796, 
	0.1866 Jy in the 60$\mu$-band was detected from its 
	companion MCG~+07-31-049.   \hfil\break
$^{\ninept 2}$\ Version~1 of the IRAS Faint Source Catalog reported marginal detection in one band only; 
	FSC2 does not report a detection in any band.\hfil\break
$^{\ninept 3}$\ No IRAS flux is reported for NGC~4174, but significant flux in all for
	IRAS bands are reported for one member of its triplet, NGC~4175.\hfil\break
$^{\ninept 4}$\ NGC~6285/6286 is an interacting pair; NGC~6286 is the 
	polar-ring candidate and the apparent source of the IRAS flux
	(M.~Schmitz, private communication). \hfil\break
$^{\ninept 5}$\ No IRAS flux is reported for NGC~235, but significant flux in all for
	IRAS bands are reported for its close companion, NGC~232. \hfil\break
$^{\ninept 6}$\ ESO~296-G11 is a superposition of two galaxies; it is uncertain which one is the 
	source of the IRAS flux.\hfil\break
}}
\end{tabular}
\clearpage